\documentclass[aps,floats,floatfix,superscriptaddress,nofootinbib,showpacs]{revtex4}
\usepackage{graphicx,epsf}
\newcommand{\e}{\rm e}

\newcommand{\be}{\begin{equation}}
\newcommand{\ee}{\end{equation}}
\newcommand{\ltil}{\widetilde\ell}

\begin{document}
\title{Condensation Transition in Polydisperse Hard Rods}

\author{M. R. Evans}\email{m.evans@ed.ac.uk}
\affiliation{SUPA, School of Physics and Astronomy, University of Edinburgh, 
Mayfield Road, Edinburgh EH9 3JZ, UK}
\author{S. N. Majumdar}\email{satya.majumdar@u-psud.fr}
\affiliation{Laboratoire de Physique Th\'eorique et Mod\`eles
Statistiques, CNRS UMR 8626, Universit\'e Paris-Sud, B\^at 100, 91405,
Orsay-Cedex, France}
\author{I. Pagonabarraga}\email{ipagonabarraga@ub.edu}
\affiliation{Departament de F\'\i sica Fonamental, Universitat de Barcelona,
Carrer Mart\`\i\ i Franqu\`es 1,
E-08028 Barcelona}
\author{E. Trizac}\email{trizac@lptms.u-psud.fr}
\affiliation{Laboratoire de Physique Th\'eorique et Mod\`eles
Statistiques, CNRS UMR 8626, Universit\'e Paris-Sud, B\^at 100, 91405,
Orsay-Cedex, France}
\email{satya.majumdar@u-psud.fr,
ipagonabarraga@ub.edu,
trizac@lptms.u-psud.fr}

\begin{abstract}
We study a mass transport model, where spherical particles
diffusing on a ring can stochastically exchange
volume $v$, with the constraint of a fixed total volume $V=\sum_{i=1}^N
v_i$,
$N$ being the total number of particles.
The particles, referred to as $p$-spheres, have a linear size that behaves
as $v_i^{1/p}$ and our model thus represents a gas of polydisperse
hard rods with variable diameters $v_i^{1/p}$.
We show that our model admits a factorized steady state distribution
which provides the size distribution that minimizes the free
energy of a polydisperse hard rod system, under the constraints
of fixed $N$ and $V$. Complementary approaches (explicit construction
of the steady state distribution on the one hand ; density functional
theory on the other hand) completely and consistently
specify the behaviour of the
system. A real space condensation transition is shown to take place
for $p>1$:  beyond a
critical density  a macroscopic aggregate is formed and
coexists with a critical fluid phase. Our work establishes the
bridge between stochastic mass transport approaches and
the optimal polydispersity of hard sphere fluids studied in previous
articles.

\end{abstract}

\maketitle

\section{Introduction}
Condensation phenomena in stochastic models of mass transport is a
subject of much current interest, for
reviews see \cite{EH05,maj09}.  Typically, in these models a
globally conserved quantity, which for the purposes of this introduction
we refer to as  mass, is transferred
stochastically between sites of a lattice according to some prescribed
dynamical rules.  One is interested in the properties of the
stationary state generated by the stochastic dynamics, for example the
single-site mass distribution which is the probability distribution of
the amount of mass at a lattice site.  Such models provide both
microscopic and effective descriptions of the dynamics of various
complex systems, for example traffic flow \cite{OEC,KMH}, granular
clustering \cite{granular}, phase ordering \cite{KLMST}, network
rewiring \cite{AELM,AHE06}, force propagation \cite{CLMNW}, aggregation and
fragmentation \cite{MKB,RM} and energy transport \cite{Bertin}. In this
paper we study  
another realisation of
stochastic mass transport in a new context: the sampling of
polydispersity in simulations of hard-rod systems.

Condensation occurs in stochastic mass transport models,
when above a critical value of the global density
a finite fraction of the mass condenses onto a single site
\cite{BBJ,OEC,MKB,Evans00}.  In some ways the phenomenon is similar in
character
to Bose-Einstein Condensation, but in contrast to Bose-Einstein
Condensation, it occurs in real space i.e. at a lattice site.  The
signature of condensation is seen in the steady-state single-site mass
distribution.  Below the critical density the distribution typically
decays exponentially for large mass, indicating a fluid phase.  At the
critical density the decay of the single-site mass distribution
is slower, typically it decays as a power law or sometimes a
stretched exponential distribution, indicating a critical fluid
\cite{MEZ05,EMZ06}. Above the critical density a bump in single-site mass
distribution emerges and corresponds to a single site containing the
excess mass above the critical value.  Thus the condensed phase
consists of  a condensate co-existing with the  critical fluid.

In the present work we consider the stochastic dynamics of mass
transfer in a rather different context, that of sampling
polydispersity in a hard rod fluid.  Polydispersity arises naturally
in a wide variety of natural and synthetic materials. In some cases,
polydispersity is an intrinsic property of the system; the elementary
constituents are characterized by a varying physical property (size or
charge, for example) which does not change as a result of the
particles' interactions. A different type of polydispersity arises as
a result of the interactions among constituent elements in situations
where they self-assemble to form aggregates of varying size, for
example.  In this second situation, the final size distribution
emerges as a result of the interaction among the constituent elements
and it will be controlled by the dynamical constraints and basic
symmetries underlying the kinetic
processes~\cite{BC01,CS02}. Previously, in \cite{ZBTCF99} a stochastic
algorithm has been used which allowed hard spheres to diffuse and to
exchange volume, subject to the hard core constraint. It turned out
that above a certain volume fraction small numbers of large particles
would emerge in the distribution generated by the dynamics. Within
approximate theories of the Percus-Yevick family, this was understood
as a continuous phase transition \cite{BC01}.

In this paper, we  will  show that the transition observed
in polydisperse hard spheres
can  in fact be understood as a condensation
transition arising from constraints in the configuration
space. To make the problem analytically solvable, we somewhat
simplify matters by constraining the spheres to move on a
one-dimensional ring (so that the particles effectively behave as hard
rods).  Basically there are two conserved quantities in play: the
total volume of the spheres and the linear size (length) of the
lattice.  However, a new feature which appears, compared to previous studies
of condensation in systems with two conserved quantities \cite{EH03},
is a configurational constraint which effectively couples the two conserved  quantities.
The configurational constraint results from the
hard-sphere condition and requires  that $x$, the separation of a sphere from
its neighbour must be greater than the diameter of the sphere defined
as $\ell= v^{1/p}$ where $v$ is the volume of the sphere ($x_i>\ell_i$
in the notation of Fig.  \ref{fig1}).  We retain $p$ as a parameter of
our particles which we refer to generally as $p$-spheres: for discs
$p=2$ and for spheres $p=3$.  The dynamics comprises diffusion, which
is a stochastic exchange of inter-particle distances, and in addition
there is a stochastic exchange of volume. When the volume fraction is
large the dynamics becomes constrained. For $p>1$, beyond a critical
volume fraction, condensation occurs. That is, in the stationary state
one large sphere emerges containing the excess volume and leaving the
rest of the system in the critical fluid phase.  When $p<1$ there is
no condensation but a vestige of the transition remains wherein above
some density threshold, the volume distribution develops a bump around
a finite characteristic volume, unlike the low density behaviour where
the most probable volume is 0.

We develop two complementary approaches to understand the condensation
phenomenon: one microscopic and one thermodynamic. In the former
approach we define a microscopic dynamical model for which we can
solve the equilibrium state exactly. We then use the usual machinery
of statistical mechanics to solve the model.  In the latter approach
we write down a free energy functional, and then use scaling arguments
to deduce various thermodynamic relations, and arrive at the same form
for the equilibrium distributions as with the microscopic approach.

\subsection{Summary of results}
As a guide to the reader, we first summarise how the paper is
organised and identify the main results obtained from our
calculations.  In section~\ref{sec:model} we define our model and 
determine simple microscopic dynamical rules which lead to an
equilibrium distribution of the factorised form (\ref{FSS}).  In
particular we identify rules which lead to an equilibrium with equal
probabilities for all allowed configurations (\ref{FSS1}) and in
section \ref{sec:gc} we consider the thermodynamics of this
equilibrium state.  
The  calculations are carried out within  the grand canonical
ensemble for the system 
which we define in \ref{sec:gce}
and within which 
the marginal distributions for
volume and separation of a particle $p(v)$ and $p(x)$ are determined
(\ref{pv},\ref{px}).  These distributions are expressed in terms of the two Lagrange
multipliers corresponding to the constraints of mean density $\rho$
and volume fraction $\phi$ (\ref{rhophigc}).  In section \ref{sec:u}
we introduce a single combination of these two Lagrange multipliers
which allows us to fix $\rho$ and consider the reduced volume fraction
$\phi^*$, defined in (\ref{phi*def}), as a function of a single
variable (\ref{phiu*def}).  We define the entropy (\ref{Sgc2}) in section
\ref{sec:S} and show that the entropy has a maximum as a function of
$\phi^*$.  We then analyse the two cases $p\leq 1$ and $p >1$
separately in sections \ref{sec:p<1} and \ref{sec:p>1}, showing that
for $p \leq 1$ the entropy increases to a maximal value then decreases
again as $\phi^*$ is increased, where for $p>1$, the entropy sticks to
its maximal value when $\phi^*$ is increased past its critical value
given by (\ref{phi*0}). The latter scenario is explained as a
condensation transition in \ref{sec:p>1}. In section \ref{ssec:largep}
we show how the limit of $p\to \infty$ recovers the results of the
Tonks gas.

In section \ref{sec:mc} we analyse the nature of the condensate by
working within the micro-canonical ensemble.  By invoking results for
the large deviations of sums of random variables we show that in the
condensed phase the marginal distribution for the volume of a particle
$p(v)$ has two pieces (\ref{pcrit},\ref{pcond}). The first piece
represents a critical fluid distribution and the second represents a
Gaussian peak corresponding to the condensate.

Having shown that our model, initially defined through a set of dynamical
rules, admits a steady state that is in fact the equilibrium state
of a hard rod system, we are then in section \ref{sec:dft} in position 
to approach the problem by a
complementary free energy functional.
We show that beginning  from the form (\ref{eq:functional})
for the free energy as a functional of the particle size disrtibution,
simple considerations imply that
the distribution of the particle diameters should follow (\ref{eq:pdfW}).
We then revisit the condensation transition in section \ref{sec:condrev}
and show the results of section \ref{sec:p>1} may be recovered.
In
section~\ref{sec:sim} we present numerical results, based on Monte Carlo simulations of the microscopic model,
confirming our
analytical predictions. In particular,
Figure \ref{fig:condensate}
confirms the emergence of a condensate at  the critical volume fraction,
and Figures  \ref{fig:aggrdist}, \ref{fig:p1.5} confirm the predictions
(\ref{pcrit},\ref{pcond})  of section~\ref{sec:mc}
for the particle volume distribution.
We conclude with an overview  in
section~\ref{sec:conc}.

\section{Model}
\label{sec:model}
In this work we introduce a simple one-dimensional model of a fluid of
hard $p$-spheres with stochastic dynamics comprising diffusion and
volume exchange.  We show that appropriate choices of the dynamical
rates allow one to obtain the steady state exactly as a factorised
form, in particular the steady state may be such that any allowed
microscopic configuration appears with equal probability.

\begin{figure}[htb]
\begin{center}
\includegraphics[width=12cm]{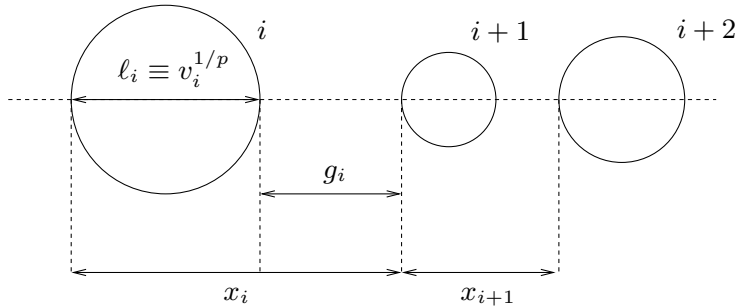}
\caption{\label{fig1}
Schematic view of three $p$-spheres on a line, with definition of the
left to left inter-particle distance $x_i$ and gap $g_i$.
}
\end{center}
\end{figure}

The model consists of  $p$-spheres diffusing on a one dimensional
ring  and exchanging volume
with hard core interactions (see Figure~\ref{fig1}).
The distance between the left hand side  of  $p$-sphere (particle) $i$
and $p$-sphere  $i+1$ is given by $x_i$ and the diameter of $p$-sphere $i$
is given by $\ell_i = v_i^{1/p}$ where $v_i$ is the volume of $p$-sphere $i$.
The volume of the  $p$-sphere with unit diameter has been set to 1
but one could easily generalise to  the volume of the unit diameter
 $p$-sphere being
$a$.

The model  is equivalent to $N$ sites $i=1,\ldots,N$ each with
site variables $x_i,v_i$ and  with periodic boundary conditions
$x_{N+1} = x_1$, $v_{N+1} = v_1$.
In the micro-canonical ensemble we have
fixed total volume
\begin{equation}
\sum_i v_i =V
\end{equation}
 and fixed total length $L$
\begin{equation}
\sum_i x_i = L\;.
\end{equation}
The
hard-core interaction between $p$-spheres implies that
for each separation $x_i$ we have the constraint
\begin{equation}
\label{con}
x_i \geq v_i^{1/p}\;.
\end{equation}
The model then can be interpreted as a system of $N$ polydisperse hard
rods on a ring where each rod has a variable diameter $\ell_i=
v_i^{1/p}$ and satisfies the hard rod constraint, namely the gap
between successive rods $g_i\ge 0$ for each $i$.  This is a
generalization of the classical monodisperse hard rod system, the
Tonks gas~\cite{Tonks}, where each rod has the same diameter $\ell$
independent of $i$. Note that our polydisperse model reduces to the
monodisperse case in the large $p$ limit since $\ell_i=v_i^{1/p}\to 1$
(independent of $i$) in the $p\to \infty$ limit.

\subsection{Dynamics}
We consider  diffusion and volume dynamics implemented by
microscopic transition rates on the variables $\{ x_i, v_i \}$.
The diffusion dynamics implies dynamics for the separations $\{ x_i \}$.
We assume symmetric nearest neighbour hopping of particles with rate
$u( \mu, x_i)$
which implies  the following exchange dynamics
for $\{ x_i \}$:
in time interval ${\rm d} t$
a length  $\mu$ is transferred from $x_i$ to $x_{i+1}$
with probability
$u( \mu, x_i) {\rm d}t/2$
and from $x_i$ to $x_{i-1}$
with probability
$u( \mu, x_i) {\rm d}t/2$. The transition is accepted if the constraints
(\ref{con}) are obeyed by the updated  variables $x_i$.

Similarly, we define the  nearest neighbour symmetric volume exchange
dynamics
with rate
$w( \Delta, v_i)$:
in time interval ${\rm d} t$
a volume  $\Delta$ is transferred from $v_i$ to $v_{i+1}$
with probability
$w( \Delta, v_i) {\rm d}t/2$
and from $v_i$ to $v_{i-1}$
with probability
$w( \Delta, v_i) {\rm d}t/2$.

\subsection{Factorised Steady State}
In order to obtain a solvable case that may allow us to study possible
condensation scenarios, we seek a factorised steady state \cite{EMZ04} where the
probabilities of a microscopic configuration $\{x_i, v_i \}$ are of the simple form
\begin{equation}
P( x_1,v_1,\ldots x_N,v_N)
= A \left[ \prod_{i=1}^N g(x_i,v_i)
\theta( x_i - v_i^{1/p})\right]
 \delta(L - \sum_i x_i) \delta(V - \sum_i v_i)
\label{FSS}
\end{equation}
where $g(x_i,v_i)$ are single-particle weights and
$A$ is a normalizing constant. 
That is, the  probability  of a configuration factorises into a product of 
one factor $g(x_i,v_i)$ for each particle.
The $\theta$-functions impose the constraint
(\ref{con}) at each site and the  $\delta$-functions impose the global
constraints of length conservation and volume conservation.
We now assume that the single site weight $g$ itself factorises
\begin{equation}
g(x_i,v_i) = a(x_i) b(v_i)\;.
\label{gab}
\end{equation}
A sufficient condition for the stationary state to be of the
form (\ref{FSS}) with $g$ of the from (\ref{gab}), is that the
$x$ and $v$ dynamics independently respect  detailed balance
with respect to $a(x)$ and $b(v)$.
Note that the constraint (\ref{con}) will  not enter into this
requirement of detailed balance
since we demand detailed balance between any two of the configurations
{\em allowed}
by the constraint.
Therefore, we require $\forall \mu,x,x',\Delta,v,v'$
\begin{eqnarray}
u(\mu,x) a(x) a(x')& =&
u(\mu,x'+ \mu) a(x-\mu) a(x'+\mu) \\
w(\Delta,v) b(v) b(v')& =&
w(\Delta,v'+ \Delta) b(v-\Delta) b(v'+\Delta)
\end{eqnarray}
which imply that $u(\mu,x)a(x)/a(x-\mu)$ is independent of $x$
and $w(\Delta,v)b(v)/b(v-\Delta)$ is independent of $v$,
leading to
\begin{eqnarray}
u(\mu,x) & =&
c(\mu) \frac{a(x-\mu)}{a(x)} \label{u} \\
w(\Delta,v) &=&
d(\Delta) \frac{b(v-\Delta)}{b(v)} \label{w}
\end{eqnarray}
where $c(\mu)$ and $d(\Delta)$ are arbitrary positive functions.
Therefore rates of the form (\ref{u},\ref{w}) lead to an equilibrium
state of the factorised form (\ref{FSS}) with single particle weights
(\ref{gab}).
Moreover,  appropriate choice
of the rates (\ref{u},\ref{w}) allow any single-site weight  (\ref{gab})
to be generated.

In the following, for simplicity  we restrict ourselves
to $g(x_i,v_i)=1$.
This requires
$u(\mu,x) = c(\mu)$ and $w(\Delta,v)= d(\Delta)$,
in order that $a(x) = b(v) =1$.
In this case  the jump size $\mu$ of a $p$-sphere does not depend on the
separation $x_i$
and $\Delta$, the amount of volume transferred from one $p$-sphere,
 does not
depend on the volume $v_i$ of the $p$-sphere (up to the constraint that
the remaining volume is non-negative).
The condition $g(x_i,v_i)=1$ implies that all allowed microscopic
configurations
have the same steady state probability. Thus the thermodynamics of the
system are driven entirely by the constraints in the configuration space  
coming from the hard-sphere condition
i.e. any phase transition will be entropy-driven.
One can easily generalize our calculation for arbitrary $b(v)$. It
turns out that the condensation transition occurs for other choices of
$b(v)$ as well, as long as the function $b(v)$ decays exponentially or
slower for large $v$. However, we stick to the choice $b(v)=1$ in this
paper for simplicity.

\section{Thermodynamics in Grand Canonical Ensemble}
\label{sec:gc}
\subsection{Ensembles}\label{sec:gce}
The natural ensemble generated by the dynamics discussed in the
previous section is the micro-canonical ensemble wherein only microscopic
configurations with the correct total length and total volume are
allowed and all allowed
configurations have the same statistical weight in the steady state
given by
\begin{equation}
P\left(\{x_i,v_i\}\right)= \frac{1}{Z_N(L,V)}\, \left[\prod_{i=1}^N\,
\theta( x_i -
v_i^{1/p})\right]\, \delta(L - \sum_i x_i)\, \delta(V - \sum_i v_i)
\label{FSS1}
\end{equation}
where $L$ and $V$ are given and the normalizing constant $Z_N(L,V)$
is just the micro-canonical partition function for $N$ particles on a ring
given by the integral
over the allowed microscopic configurations
\begin{equation}
Z_N(L,V)=  \prod_{i=1}^N \int {\rm d}x_i\, {\rm d}v_i\,
\theta(x_i- v_i^{1/p})\,   \delta(L -
\sum_i x_i)\, \delta(V - \sum_i v_i)\;,
\label{Zcdef}
\end{equation}
i.e. $Z_N(L,V)$ is a volume in configuration space.
We define the two basic control parameters: the density of particles
$\rho=N/L$ and the
volume per particle $\phi=V/N$. Our goal in this section is to compute,
in the
large $N$ limit but for
given fixed $(\rho,\phi)$, the single-site distribution $p(x,v)$
obtained
from the joint distribution (\ref{FSS1}) by integrating out all the
$\{x_i,v_i\}$
variables except at one site where they are held fixed with values
$(x,v)$. Next,
from this single-site distribution $p(x,v)$ we will derive the marginals
$p(x)=\int_0^{\infty}
p(x,v)\, dv$ and $p(v)= \int_0^{\infty} p(x,v)\, dx$ for given
$(\rho,\phi)$.
We will show that the marginals $p(x)$ and $p(v)$ exhibit a rich variety
of behavior in different regions of the $(\rho,\phi)$ plane, including
a condensation transition for $p>1$.
 
To make progress, we shall first follow the usual route of replacing
the hard constraints on the allowed total length and total volume in
the micro-canonical ensemble by the soft constraints of the grand
canonical ensemble where the total length and volume are allowed to
fluctuate.  Thus, for this system we define the grand canonical
partition function as the double Laplace transform of the
micro-canonical partition function $Z_N(L,V)$
\begin{equation}
{\cal Z}_N(\lambda,s)=
\int_0^\infty {\rm d}L \int_0^\infty{\rm d}V Z_N(L,V) {\e}^{-L\lambda -Vs}\;.
\label{Zgcdef}
\end{equation}
Then inserting (\ref{Zcdef}) into (\ref{Zgcdef}) yields
\begin{eqnarray}
{\cal Z}_N(\lambda,s)&=&
 \prod_{i=1}^N \int {\rm d}x_i {\rm d}v_i
\theta(x_i- v_i^{1/p})\ {\e}^{-\lambda \sum_ix_i -s \sum_i v_i}
\\
&=& \prod_{i=1}^N \int_0^\infty {\rm d}v_i
\prod_{i=1}^N \left[ \int_{v_i^{1/p} }^{\infty} {\rm d}x _i
\ {\e}^{-\lambda x_i -s  v_i}\right] \\
&=&
\left[ G(\lambda,s)\right]^N\;,
\label{Zgc}
\end{eqnarray}
where  the single-particle  partition function $G(\lambda,s)$ is defined as
\begin{equation}
G(\lambda,s) = \frac{1}{\lambda}
\int_0^\infty {\rm d}v
 {\e}^{-sv -\lambda v^{1/p} }\;.
\label{gslambda}
\end{equation}
Thus in the grand canonical ensemble the joint distribution (\ref{FSS1})
factorises $P(\{x_i,v_i\})=\prod_i p(x_i,v_i)$ with the single-particle
joint distribution $p(x,v)$ given by
\begin{equation}
p(x,v) = \left[G(\lambda,s)\right]^{-1} {\e}^{-\lambda x -s  v}
\theta(x-v^{1/p}) \theta(v)\;,
\label{pxv}
\end{equation}
where we have made explicit the fact that $v\geq 0$.
One can integrate out $x$ or $v$ to obtain respectively,
the single-particle volume and length  marginal distributions
\begin{eqnarray}
p(v) &=& \left[G(\lambda,s)\right]^{-1}\frac{1}{\lambda}
{\e}^{-\lambda v^{1/p} -s  v}\theta (v) \label{pv}\\
p(x) &=& \left[G(\lambda,s)\right]^{-1}\frac{1}{s}  {\e}^{-\lambda x}
\left[ 1- {\e}^{ -s  x^p}\right] \theta(x)\;. \label{px}
\end{eqnarray}
The constraint $\theta(x-v^{1/p})$ in (\ref{pxv}) makes the
$x$ and $v$ variables manifestly coupled. Note however that
$p(x,v)$ can be `diagonalised' or decoupled
if one uses the gap variables
$g_i = x_i -v_i^{1/p}$ (see Fig. \ref{fig1}) instead of the
$x_i$ variables.
Then $p(x,v)= {\tilde p}(g,v)$ becomes
\begin{equation}
{\tilde p}(g,v) = \left[G(\lambda,s)\right]^{-1}\, \left[e^{-\lambda\,
g}\,\theta(g)\right]\,\left[e^{-sv-\lambda v^{1/p}} \theta(v) \right].
\label{diag1}
\end{equation}
The consequence of this decoupling will be discussed later.

It remains then to estimate the two Lagrange multipliers $(\lambda,s)$
by enforcing the conservation of total length and total volume
on an average.
The mean lattice length $\overline{L}$ and mean volume
$\overline{V}$  in the grand canonical ensemble are given by
\begin{equation}
\overline{L} = -\frac{ \partial \ln{\cal  Z}_N(\lambda,s)}{\partial
  \lambda}\;;
\qquad
\overline{V} =  -\frac{ \partial \ln{\cal Z}_N(\lambda,s)}{\partial s}\;.
\label{gc}
\end{equation}
In this ensemble, our original control parameters $(\rho,\phi)$ are thus
replaced by the averages
\begin{eqnarray}
\rho =   \frac{N}{\overline{L}}\label{rhodef}\;;\qquad
\phi= \frac{\overline{V}}{N}\;.
\end{eqnarray}
Using the definitions (\ref{gc}) and the expression of $\cal Z_N$
 (\ref{Zgc})
gives
\begin{equation}
\frac{1}{\rho} = - \frac{\partial \ln G(\lambda,s)}{\partial \lambda}\;;
\qquad \phi  = - \frac{\partial \ln G(\lambda,s)}{\partial s}\;.
\label{rhophigc}
\end{equation}
Thus, for given values of the control parameters $(\rho,\phi)$, we
have to solve the two conditions (\ref{rhophigc}) 
to get the corresponding values
$(\lambda,s)$ and then use them in (\ref{pv}) and (\ref{px}) to obtain
the marginals within the grand canonical framework.

For the discussion in Section \ref{sec:dft} on the alternative free energy
functional route, it turns out to be convenient to introduce another
dimensionless observable denoting the line coverage
\begin{equation}
\eta= \frac{L_p}{\overline L}
\label{etadef}
\end{equation}
where $L_p = N \langle v_i^{1/p} \rangle$ is
the average total length occupied by the $p$-spheres.
In other words, $\eta$ is the ratio of $\langle v_i^{1/p} \rangle$, the
average length of a
$p$-sphere, to $1/\rho$, the average available length.
The first average is  also easy to compute within the grand canonical
ensemble
using (\ref{gslambda})
\begin{equation}
\langle v_i^{1/p} \rangle= \frac{ \int_0^{\infty} v^{1/p}\, {\e}^{-sv
-\lambda v^{1/p}}\,
{\rm d} v}{\lambda G(\lambda,s)}=
-\frac{\partial}{\partial
\lambda}\ln \left[\lambda G(\lambda,s)\right].
\end{equation}
The right hand side can be expressed in terms of the density
(\ref{rhophigc}) and we get
\begin{equation}
\lambda= \frac{\rho}{1-\eta}.
\label{etadef1}
\end{equation}
The quantity on the right hand side
of (\ref{etadef1}) is nothing but the pressure
of a polydisperse hard-rod fluid \cite{TP08}. As expected, we find that
the Lagrange multiplier $\lambda$ associated to the conservation of total
length $L$ coincides with the pressure (see section~\ref{sec:opd})

Substituting $\lambda$ from (\ref{etadef1}) in (\ref{diag1}), we find that
the marginal gap distribution ${\tilde p}(g)= \int_0^{\infty} {\tilde
p}(g,v)\, {\rm d}v$
for this polydisperse hard-rod fluid is given by
\begin{equation}
{\tilde p}(g) = \frac{\rho}{1-\eta}\,
\exp\left[-\frac{\rho\,g}{1-\eta}\right]\,\theta(g).
\label{diag2}
\end{equation}
We note that in the case of monodisperse hard rods
(the Tonks gas, where all rods have the same length)
the equilibrium  gap distribution has exactly the same form as
(\ref{diag2})~\cite{Tonks}. Here we see that expression (\ref{diag2})
is more general and holds even for the gap distribution of a polydisperse
hard-rod fluid.

\subsection{Reduced volume fraction $\phi^*$ and scaling variable $u$}
\label{sec:u}
To facilitate further analysis it is useful to
consider a scaling combination of the two Lagrange multipliers
$s$, $\lambda$
\begin{equation}
u=\frac{s}{\lambda^p}\;.
\label{udef}
\end{equation}
If we make a change of variable $v=\lambda^{-p}\,y$
the single-particle partition function $G(\lambda,s)$ in
(\ref{gslambda}) can
be written in the scaling form
\begin{equation}
G(\lambda,s)=
\frac{1}{\lambda^{p+1}}\,H\left(\frac{s}{\lambda^p}\right)
\label{G(u)}
\end{equation}
where
\begin{equation}
H(u)\equiv
\int_0^{\infty} {\rm d}y {\e}^{-uy-y^{1/p}}.
\label{Hdef}
\label{gslambdasc1}
\end{equation}

Similarly, the  constitutive equations for the density and volume fraction (\ref{rhophigc}) may be cast in a scaling form in terms of  the  variable  $u$
\begin{eqnarray}
\frac{1}{\rho} &= &\frac{1}{\lambda}\,
F_\rho\left(\frac{s}{\lambda^p}\right)
\label{rhoscaling1} \\
\phi &=& \frac{1}{\lambda^p}\, F_{\phi}\left(\frac{s}{\lambda^p}\right)
\label{phiscaling1}
\end{eqnarray}
where the two scaling functions $F_\rho(u)$ and $F_\phi(u)$ are given by
\begin{eqnarray}
F_{\rho}(u) & =& 1 + \frac{\int_0^{\infty} {\rm d}y\, y^{1/p}\,
{\e}^{-uy-y^{1/p}}}{\int_0^{\infty}
{\rm d}y\, {\e}^{-uy-y^{1/p}}} \label{rhoscaling2}\;,\\
F_{\phi}(u) &=& \frac{\int_0^{\infty} {\rm d}y\, y\,
{\e}^{-uy-y^{1/p}}}{\int_0^{\infty}
{\rm d}y\, {\e}^{-uy-y^{1/p}}} = - \frac{ {\rm d} \ln H(u)}{ {\rm d} u}\;.
\label{phiscaling2}
\end{eqnarray}
As a result, we can eliminate $\lambda$ between (\ref{rhoscaling1}) and
(\ref{phiscaling1}) by introducing the scaled volume per particle, $\phi^*$,
defined by
\begin{equation}
\phi^* = \phi^{1/p}\rho\,
\label{phi*def}
\end{equation}
which depends only on a single scaled variable $u=s/\lambda^p$:
\begin{equation}
\phi^*(u) = \frac{[F_{\phi}(u)]^{1/p}}{F_{\rho}(u)}\;.
\label{phiu*def}
\end{equation}
We call $\phi^*$  the {\em reduced volume fraction}.
Such a reduction to a single scaling variable (instead
of two independent variables $(\rho,\phi)$) can be traced back to the fact that
in the grand canonical ensemble the joint distribution $p(x,v)$, when
expressed in terms of the gap $g_i$ variables, essentially decouples
as in (\ref{diag1}).
In due course we will investigate the behavior of the function
$\phi^*(u)$ for different values of $p$.

\subsection{Entropy}\label{sec:S}
As already stated the condensation transition in our model
is  completely entropy driven, therefore we  should first
define the entropy. In terms of the
original micro-canonical partition function (\ref{Zcdef}), the 
entropy per particle is
\begin{equation}
S\equiv \frac{\ln Z_N(L,V)}{N}.
\label{entropy1}
\end{equation}
Since the configuration space volume $Z_N(L,V)$ (\ref{Zcdef}) may be
less than 1,
$S$, as defined above, has no reason to be a positive definite variable.
The entropy could of course  be made positive by  dividing $Z_N$ by a
suitable small constant  in the same way that the partition function for an
ideal gas contains a factor $h^{-3N}$.

In the grand canonical ensemble, the entropy per particle may be expressed
as the Legendre transform %
\begin{eqnarray}
S& \equiv & \frac{1}{N} \left[ \ln {\cal Z}_N(\lambda, s)
- \lambda \frac{\partial \ln {\cal Z}_N(\lambda, s)}{\partial \lambda}
- s\frac{\partial  \ln {\cal Z}_N(\lambda, s)}{\partial s} \right]\\
\label{Sgc1}
&=& \ln G(\lambda,s) + \frac{\lambda}{\rho} + s \phi \;,
\label{Sgc2}
\end{eqnarray}
where in the second line we impose the constraints (\ref{rhophigc}).
A consequence of  (\ref{Sgc2}) is that
\begin{equation}
\frac{\partial S}{\partial \phi}\Big|_{\rho} = s\;;
\qquad
-\rho^2 \frac{\partial S}{\partial \rho}\Big|_{\phi} = \lambda\;.
\label{entropyder2}
\end{equation}
In particular, we note that the derivative of $S$ with respect to $\phi$
vanishes at $s=0$.

We now consider the entropy as a function of the scaling variable
$u=s/\lambda^p$, introduced in Eq. (\ref{udef}).
For fixed density $\rho$, one can  express the entropy $S$ as a single
function of $u$.
Substituting (\ref{rhoscaling1}),
(\ref{phiscaling1}) and (\ref{G(u)}) in (\ref{Sgc2}) allows us
to write
\begin{equation}
S = -(p+1)\,\ln \rho -(p+1)\,
\ln F_{\rho}(u)+\ln H(u) +F_{\rho}(u)+u\,F_{\phi}(u).
\label{Su}
\end{equation}
For fixed density $\rho$, as $\phi$ and hence $\phi^*=\phi^{1/p}\rho$ varies, $u$ varies according to
(\ref{phiu*def})
and consequently %
the entropy $S$ varies.

\begin{figure}[htb]
\begin{center}
\includegraphics[width=9cm]{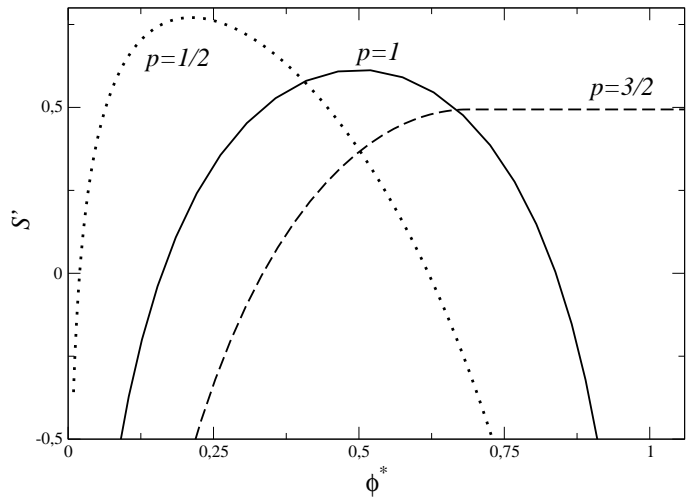}
\caption{\label{fig:entropy}
Entropy plot of $S' \equiv S + (p+1) \log \rho$,
where $S$ is the grand canonical  entropy,
as a function of reduced density $\phi^*$, for different values
of parameter $p$. The three corresponding critical densities are
$\phi^*_c = 2/(3 \pi) \simeq 0.212$ for $p=1/2$, $\phi^*=1/2$ for $p=1$ and
$\phi^*_c=4 (2/\pi)^{1/3}/5 \simeq 0.688$ for $p=3/2$. In the first two
cases
no transition occurs. For $p>1$, the plateau seen
for $\phi^*>\phi_c^*$ is a signature of the condensate formation.
Note that for $p\leq 1$, the maximum allowed value of $\phi^*$ is 1,
whereas one can have $\phi^*>1$ for $p>1$.
}
\end{center}
\end{figure}

In Appendix \ref{app:A}, we show from expression (\ref{Su}) that the entropy
achieves a maximum when $u=0$.  With this knowledge we will
investigate in the next two subsections how the entropy $S$, for a
fixed density $\rho$, behaves as a function of $\phi^*$ for $p\le 1$ and
for $p>1$.  The results are shown in Figure~\ref{fig:entropy}.
We will see that a condensation transition occurs only for
$p>1$  where, when $\phi^*$ is increased above a critical value, $u$
sticks to the value $u=0$ and the entropy sticks to its maximal value.
A vestige of the condensation transition can also be seen for $p\le 1$
in the fact that the entropy decreases when $\phi^*$ exceeds the
critical value (and $u$ passes through zero) which reflects that the
configuration space becomes severely constrained.

\null\vskip 1em

\subsection{The case $p\le 1$}\label{sec:p<1}
Recall that, according to our model, for $p < 1$ the volume of the
$p$-sphere increases sub-linearly with its diameter and for $p = 1$ the
volume is equal to the diameter.  Let us start by making the following
observation. By definition, the inverse density is $1/{\rho}= L/N =
(\sum_i x_i)/N$. Due to the constraint $x_i\ge v_i^{1/p}$, it follows
that $1/{\rho}\ge (\sum_i v_i^{1/p})/N$. For $p\le 1$, one can use
Jensen's inequality to write
\begin{equation}
\frac{1}{\rho}\ge \frac{\sum_i v_i^{1/p}}N \ge \left[\frac{\sum_i
v_i}{N}\right]^{1/p}=\phi^{1/p}.
\label{Jensen1}
\end{equation}
Thus, for $p\le 1$, one must necessarily have $\phi^* = \phi^{1/p}\rho
\le 1$. In other
words, for a given $\rho$, one is physically allowed to increase $\phi$
only up to
$\rho^{-p}$, i.e., $\phi\le \rho^{-p}$.

Now, let us fix the density $\rho$ and imagine increasing $\phi$ from
$0$ to its maximally allowed value $\rho^{-p}$. In other words, the
reduced volume fraction $\phi^* = \phi^{1/p} \rho$ increases from $0$
to its maximally allowed value $1$.  For a given $\phi^*$, we then
have to solve (\ref{phiu*def}) for $u=s/\lambda^p$.\\

\begin{figure}[htb]
\begin{center}
\includegraphics[width=8cm]{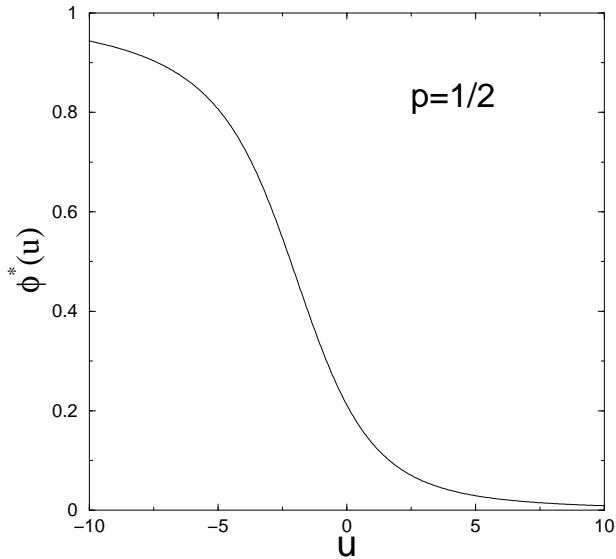}
\caption{\label{fig:phiu}
The function $\phi^*(u)$ vs $u$ for $p=1/2$.
}
\end{center}
\end{figure}

\noindent{{\em The case} $\bf{p<1}$:}
Consider first the case when $p<1$ strictly. The case $p=1$ will be
discussed subsequently.
For $p < 1$, we notice that the integrals in (\ref{rhoscaling2}) and
(\ref{phiscaling2})
are convergent for any $u\in [-\infty,\infty]$.
Their leading asymptotic
behaviors can
be easily deduced. Specifically, one finds that as $u\to \infty$
\begin{eqnarray}
F_{\rho}(u) &\to & 1 \label{rhoasympr} \\
F_{\phi}(u) &\to & \frac{1}{u}. \label{phiasympr}
\end{eqnarray}
On the other hand, as $u\to -\infty$, to leading order,
\begin{eqnarray}
F_{\rho}(u) &\to & |u|^{1/(1-p)} \label{rhoasympl} \\
F_{\phi}(u) &\to & |u|^{p/(1-p)}. \label{phiasympl}
\end{eqnarray}
As a consequence, the function $\phi^*(u)$ in (\ref{phiu*def}) has the
following
asymptotic behavior
\begin{eqnarray}
\phi^*(u) &\to & 1 \quad {\rm as}\quad u\to -\infty  \\
& \to & u^{-1/p} \quad {\rm as}\quad u\to \infty.
\label{phi*asymp}
\end{eqnarray}
In addition, one can check that $\phi^*(u)$ is a monotonically decreasing
function of $u$, achieving its maximally allowed value $1$ as $u\to
-\infty$ (see Fig. \ref{fig:phiu}).
Thus, for any given $\phi^*$, we can always find a solution $u$
to the equation $\phi^*= \phi^*(u)$ where $\phi^*(u)$ is given in
(\ref{phiu*def}).
Knowing this solution $u$, one finds subsequently $\lambda>0$ from
(\ref{rhoscaling1}) and $s$ from the relation $s= u\,\lambda^p$.
This means that the grand canonical framework works over the
full allowed range $0\le \phi^*\le 1$ and the two marginals $p(v)$
and $p(x)$ have always the form in (\ref{pv}) and (\ref{px}) with
$\lambda$ and $s$ determined as above. {\it This shows that there is
no condensation for} $p< 1$.

However, a vestige of the condensation transition still remains even
for $p< 1$. To see this, imagine again that we increase the value of
the control parameter $\phi^*=\phi^{1/p}\rho$ from $0$ to $1$. As
$\phi^*$ is increased, the solution $u$ to the equation
(\ref{phiu*def}) decreases monotonically from $\infty$ to $-\infty$ (see Fig. \ref{fig:phiu}).
Note that when the solution $u$ hits $0$, the corresponding value
of $\phi^*(0)$ can be computed by putting $u=0$
in (\ref{phiu*def}) and carrying out the elementary integrals
that gives $F_{\rho}(0)= (p+1)$ and $F_{\phi}(0)= \Gamma(2p)/\Gamma(p)$.
Thus
\begin{equation}
\phi^*(0)= \frac{1}{1+p}\,
\left[\frac{\Gamma(2p)}{\Gamma(p)}\right]^{1/p} \;.
\label{phi*0}
\end{equation}

Now, let us investigate the entropy $S$ in (\ref{Sgc2}) as a function
of the control parameter $\phi^*$.  As $\phi^*$ is increased monotonically from $0$
(and consequently $u$ decreases monotonically from
$+\infty$), one can check that the entropy $S$, expressed as a
function of $u$ as in (\ref{Su}), increases monotonically as long as
$u>0$, i.e., $\phi^*< \phi^*(0)$ given in (\ref{phi*0}). For $\phi^*>
\phi^*(0)$ (or $u<0$), the entropy decreases with increasing
$\phi^*$. Thus, the entropy $S$ has a maximum value at $\phi^*=
\phi^*(0)$, see Fig. \ref{fig:entropy}.  This is also evident from
(\ref{entropyder2}) where the derivative of the entropy as a function
of $\phi$ (for fixed $\rho$) vanishes at $s=0$ and hence at
$u=0$. Thus the value of $\phi$ at which the entropy becomes a maximum
(for a fixed $\rho$) can be appropriately denoted by
$\phi_{max}(\rho)$ and is given by
\begin{equation}
\phi_{max}(\rho)= \left[\frac{\phi^*(0)}{\rho}\right]^p =
\frac{1}{\rho^p}\,\frac{1}{(1+p)^{p}}\, \frac{\Gamma(2p)}{\Gamma(p)}\;.
\label{phimaxp<1}
\end{equation}
The corresponding maximum value of the entropy
is obtained
by putting $u=0$ in (\ref{Su}). Using $F_{\rho}(0)=(p+1)$ and
$H(0)=\Gamma(p+1)$ from (\ref{G(u)}) we get
\begin{equation}
S_{max}= p+1+\ln \Gamma(p+1) -(p+1)\, \ln ((p+1)\rho)\;.
\label{Smax1}
\end{equation}
Physically this means that for $\phi>\phi_{max}(\rho)$ or equivalently
in terms of the reduced volume fraction, for $\phi^*> \phi^*(0)$, the
configuration space becomes constrained resulting in the reduction
of entropy.

Can one see a reflection of this vestige of a condensation transition directly in the volume distribution $p(v)$ in Eq. (\ref{pv})? Indeed one does observe
a change of behavior of $p(v)$ as the control parameter $\phi^*$ increases
through $\phi_{\max}(\rho)$.
As $\phi^*$ increases from $0$,
the solution $u$ of $\phi^*(u)=u$ is positive as long as $\phi^*<\phi_{max}(\rho)$
(see Fig. \ref{fig:phiu}).
Consequently $s=u \lambda^p$ is also positive and hence the distribution 
$p(v)\propto \exp[-\lambda 
v^{1/p}-s v]$ is a monotonically decreasing function of $v$ with its maximum at $v=0$.
However, when $\phi^*> \phi_{max}(\rho)$, the solution $u$ and hence $s$ becomes negative.
As a result, $p(v)\propto \exp[-\lambda
v^{1/p}+|s| v]$ now develops a maximum at a nonzero characteristic volume
$v^*= (p|s|/\lambda)^{p/(1-p)}$.\\

\noindent{{\em The case} $\bf{p=1}$:} The general conclusion reached above
for $p<1$ remains valid even for the marginal case $p=1$, though the details
are slightly different. Indeed, for $p=1$, we can obtain explicit solutions
for the marginal distributions. To see this, let us first  
explicitly express $\lambda$ and
$s$ in terms of $\rho$ and $\phi$. The integrals in (\ref{rhoscaling2})
and (\ref{phiscaling2}) can be easily performed explicitly for $p=1$ giving
\begin{equation}
F_{\rho}(u) = \frac{u+2}{u+1}\;;\quad F_{\phi}(u) =\frac{1}{u+1}.
\label{p1phirho}
\end{equation}
Thus, unlike the $p<1$ case where the allowed range of $u$ was $u\in [-\infty,\infty]$,
for $p=1$, $u$ has the allowed range $u\in [-1,\infty]$. 
Eq. (\ref{phiu*def}) then gives
\begin{equation}
\phi^*(u) =\frac{1}{u+2}.
\label{p1phi*}
\end{equation}
Thus $\phi^*(u)$ is again a monotonically decreasing function of $u$ in $u\in [-1,\infty]$
and for any given $\phi^*$ one can always find a solution $u=1/\phi^*-2$. Hence,
as in the case $p<1$, {\it there is no condensation for $p=1$} as well.

Note that at $u=0$, $\phi^*(0)=1/2$ and hence $\phi_{max}(\rho)=1/(2\rho)$.
Also, (\ref{rhoscaling1}) and (\ref{phiscaling1}) yield explicitly
\begin{eqnarray}
\phi = \frac{1}{s+\lambda} \label{phip1}\\
\frac{1}{\rho}-\phi = \frac{1}{\lambda}\;, \label{rhop1}
\end{eqnarray}
determining $\lambda=\rho/(1-\rho \phi)$ and $s= (1-2\rho\phi)/(1-\rho
\phi)$
in terms of $\rho$  and $\phi$. This also gives, using (\ref{etadef}),
$\eta=\rho \phi=\phi^*$.
The grand canonical joint distribution
for the separation  and volume of a particle  (\ref{pxv}) is
\begin{equation}
p(x,v) = \lambda(s+\lambda) e^{-x \lambda -v s} \theta(x-v)\;.
\end{equation}
The marginal distributions (\ref{pv},\ref{px}) become
\begin{eqnarray}
p(v)& =& \frac{{\e}^{-v/\phi}}{\phi}\\
p(x)& =& \frac{\rho}{1-2\,\rho\,\phi}\left[ {\e}^{-{\rho\,
x}/(1-\rho\,\phi)} -
{\e}^{-x/\phi}\right]\;.
\end{eqnarray}

Thus, although there is no phase transition for $p=1$,
there is a change in the leading  behaviour of $p(x)$ as
the volume per particle  $\phi$ is increased past
$\phi_{max}(\rho)=1/(2\rho)$:
for $\phi < 1/(2 \rho)$ the large $x$ behaviour is
$p(x) \simeq {\e}^{-x\rho/(1-\rho\,\phi)}\rho /(1-2\,\rho\,\phi)$;
for $\phi = 1/(2\rho)$ the large $x$ behaviour is
$p(x) =  x {\e}^{-x/\phi}/\phi^2$
(note that this expression, for $\phi=1/(2\rho)$, holds for all values
of left-to-left distances $x$);
for $\phi > 1/(2\rho)$ the large $x$ behaviour is
$p(x) \simeq {\e}^{-x/\phi}\rho/(2\,\phi\,\rho -1)$.
Thus for  $\phi > 1/(2\rho)$, the exponential decay of $p(x)$ is the same
as that of $p(v)$.

The entropy per particle (\ref{Sgc2})
 reads
\begin{equation}
S=  \ln(1-\rho\phi) -\ln\rho + \ln \phi +  2.
\label{entropyp1}
\end{equation}
Note that, as mentioned below (\ref{Sgc2}), the entropy $S$ is not
restricted to be only positive. This is evident
in the $p=1$ case from (\ref{entropyp1}) where $S$ can be negative for
certain values of the parameters $\rho$ and $\phi$.
One can verify easily that the entropy $S$ in (\ref{entropyp1}) has
a maximum at $\phi=\phi_{max}(\rho)=1/(2\rho)$ with
value $S_{\max}= 2+\ln (2)-2\ln (2\rho)$ from (\ref{Smax1}),
see also Fig. \ref{fig:entropy}.
As $\phi$ increases past $1/(2\rho)$
the entropy decreases,  implying
a constrained configuration space. Thus the scenario
for $p=1$ case is similar to $p<1$ discussed earlier, as
seen clearly in Fig. 2 for the two representative cases $p=1/2$
and $p=1$.

\subsection{The case p$>$1}\label{sec:p>1}

For $p>1$, the integrals on the right hand side of (\ref{rhoscaling2})
and (\ref{phiscaling2}) are convergent only for $u>0$. Thus the
lowest allowed value of $u$ is $0$. When $u\to 0$, the function
$\phi^*(u)$ in (\ref{phiu*def}) approaches $\phi^*(0)$ which is
still given exactly by (\ref{phi*0}). Thus this is the
maximum value of $\phi^*$ allowed by the grand canonical ensemble.
If $\phi^*$ exceeds $\phi^*(0)$, $u=s/\lambda^p$ cannot decrease below $0$.
It then sticks to its value $u=0$ and all the extra volume condenses into
a single site. Thus $\phi^*(0)$ is the critical value beyond which
the grand canonical ensemble
breaks down signalling the onset of a condensation transition.

At this critical point, the volume fraction $\phi= [\phi^*(0)/\rho]^p$
will again be denoted by $\phi_{max}(\rho)$ and has the same expression as
in the $p\le 1$ case, namely
\begin{equation}
\phi_{max}(\rho)=\frac{1}{\rho^p}\,\frac{1}{(1+p)^{p}}\,
\frac{\Gamma(2p)}{\Gamma(p)}\;.
\label{eq:phimax}
\end{equation}
Also, letting $u=0$ in (\ref{rhoscaling1}) gives the critical value
$\lambda_c = (p+1)\rho$. Consequently, the two marginals in (\ref{pv})
and (\ref{px})
at the critical point become
\begin{eqnarray}
p(v) &\to& \left[G(0,\lambda)\right]^{-1}\frac{1}{\lambda}  {\e}^{-\lambda
  v^{1/p}}
= \frac{1}{\Gamma(1+p)}\,\left[\rho\,(1+p)\right]^p
{\e}^{-\rho\,(1+p)\,v^{1/p}}
\label{pvc}
\\
p(x) &\to& \left[G(0,\lambda)\right]^{-1} {\e}^{-\lambda x}  x^p
= \frac{1}{\Gamma(1+p)}\, \left[\rho\,(1+p)\right]^{1+p}\,x^p\,
{\e}^{-\rho\,(1+p)\,x}
\label{pxc}
\end{eqnarray}
Thus at condensation $p(v)$ changes from (dominant) exponential
decay (\ref{pv}) to a slower stretched exponential decay (\ref{pvc})
and $p(x)$ changes from an exponential
decay (\ref{px}) to the exponential  decay multiplied by $x^p$ (\ref{pxc}).
We interpret these results as  describing the critical fluid;
in the condensed phase we expect a condensate to coexist with the
critical fluid.
For later purposes, we also note that at the critical point
the dimensionless line coverage
\begin{equation}
\eta\to \eta_c = \frac{p}{(1+p)}
\label{etacrit}
\end{equation}
which follows by substituting $\lambda_c=(p+1)\rho$ in
(\ref{etadef1}).

How does the entropy $S$ behave as $\phi$ increases from $0$ to
$\phi_{max}(\rho)$
for fixed $\rho$? As $\phi$ increases monotonically, $u$ decreases
monotonically
till $u$ hits $0$. Consequently, the entropy $S$ in (\ref{Su}) increases
monotonically up to $u=0$, achieving a maximum value at $u=0$ (or
equivalently at
$\phi=\phi_{max}(\rho)$) given by the same expression as in the case
$p\le 1$
in (\ref{Smax1}) namely,
\begin{equation}
S_{max}= p+1+\ln \Gamma(p+1) -(p+1)\, \ln ((p+1)\rho)).
\label{Smax2}
\end{equation}

What happens to the entropy when $\phi$ exceeds the critical value
$\phi_{\max}(\rho)$,
i.e., when a condensate sets in?
To see this, we note that $\phi_{max}(\rho)$ in (\ref{eq:phimax})
can be neatly expressed in terms of the $p$-th moment $\langle x^p\rangle$
of the critical marginal $p(x)$ in (\ref{pxc}). This moment can be
easily computed and comparing to the expression of $\phi_{max}(\rho)$
in (\ref{eq:phimax}) one easily verifies that
\begin{equation}
\phi_{max}(\rho) = \frac{1}{2} \langle x^p \rangle\;.
\label{phimax}
\end{equation}
The physical meaning of (\ref{phimax}) is that
condensation occurs when the volume per particle is equal to half the mean
available volume per particle in the grand canonical distribution.
In the case $p\le 1$ this is the value of $\phi$ above which the
entropy decreases.
For $p>1$,
the intuitive
explanation is that at this point, rather than the entropy decreasing as
was
the case for $p=1$, the entropy can be held constant by one particle
containing the excess volume and leaving the rest of the system at the
critical
volume fraction, see Fig. \ref{fig:entropy}. 

\begin{figure}[htb]
\begin{center}
\includegraphics[width=7cm]{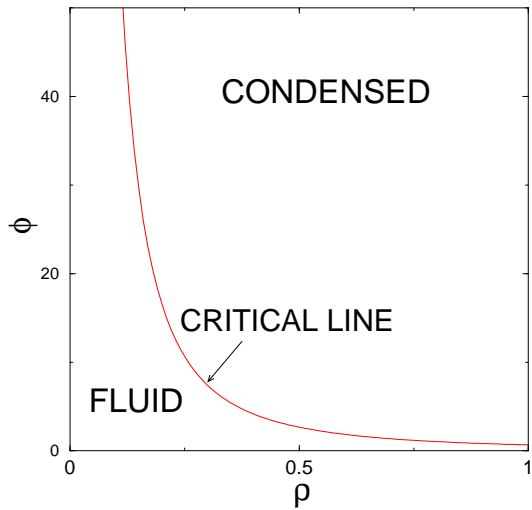}
\caption{\label{fig:phd} Phase diagram in the $(\rho,\phi)$ plane for
$p=2$.
The critical line $\phi_{max}(\rho)= \frac{2}{3\,\rho^2}$ (for $p=2$)
separates the fluid
and the condensed phase.
}
\end{center}
\end{figure}

Let us then summarize the $p>1$ case.  
The existence of the grand canonical solution means that the system
is in a fluid state. Thus, for fixed $\rho$, $\phi_{\max}(\rho)$ in
(\ref{eq:phimax})
is precisely the critical line in the $(\rho,\phi)$ phase diagram, as
shown in Fig. (\ref{fig:phd}) for $p=2$. When the volume fraction
$\phi$ exceeds this critical value $\phi_{\max}(\rho)$, the grand canonical
description breaks down and a condensate forms in the system. The entropy
$S$ increases monotonically with increasing $\phi$ till $\phi$ hits
the critical value $\phi_{max}(\rho)$ where it achieves its maximum
value $S_{max}$ given in (\ref{Smax2}). When $\phi$ is increased to  a value
$\phi > \phi_{max}(\rho)$, the entropy remains at the maximal value through one particle
containing the excess volume and forming a condensate.

\subsection{The large p limit}
\label{ssec:largep}

The limit $p\to \infty$ is worth mentioning also as it makes
links with some other well studied problems.
As mentioned in the begining of Section 2, in the
$p\to \infty$ limit our polydisperse hard-rod system
reduces to the classical monodisperse Tonks gas of hard rods where
each rod has the same size $1$. In this case, our result
for the gap distribution in (\ref{diag2}) (valid for general $p$),
reduces to that of the classical Tonks gas.
 
On the other hand (again for $p\to \infty$), the volume of each rod
becomes a ``passive'' property (a color $v \in [0,\infty[)$.  
From (\ref{pv}), this property is exponentially distributed
($p(v) \propto \exp(-s v)$). Such a Poissonian distribution may have been
anticipated, and is readily obtained within the free energy functional
formalism presented in section \ref{sec:dft}. We note that this
exactly coincides with the exponential distribution of wealth
and income obtained in the model presented in
reference \cite{Dragu}.

We also note that in the large $p$
limit, the condensation transition disappears. Thus the system is always
in a fluid state and the volume
statistics obey ``ideal gas'' behaviour. In the hard core language
of section \ref{sec:dft}, we will indeed see that the critical
line fraction tends to unity for large $p$, so that the parameter
range $\phi>\phi_{max}$ actually corresponds to an unphysical region
where ``hard rods'' necessarily overlap. Hence, the model of Ref
\cite{Dragu} shows no condensation transition (more complex interactions
between the ``agents'' would be required).

\section{Micro-canonical Analysis of the Condensed Phase for $p>1$}
\label{sec:mc}
For $p>1$ the grand canonical ensemble can only support a volume per
particle $\phi
\leq \phi_{max}(\rho)$, therefore in order
to fully analyse the condensed phase, where $\phi=V/N>
\phi_{max}(\rho)$, one needs to work
within the micro-canonical ensemble.

To compute the micro-canonical partition function
one inverts the Laplace transforms in (\ref{Zgc})
\begin{eqnarray}
Z_N(L,V) &=&  
\int_{c_1 -i \infty}^{c_1+ \infty}
\frac{{\rm d}s}{2\pi i}
\int_{c_2 -i \infty}^{c_2 + \infty}
\frac{{\rm d}\lambda}{2\pi i}
{\e}^{sV + \lambda L} {\cal Z}_N(\lambda,s)
\end{eqnarray}
where $c_1$ and $c_2$ are chosen so that
the integration contours are to the right of any singularities.
Using  the expression (\ref{Zgc}) one finds
\begin{eqnarray}
Z_N(L,V) &=&  
\int_{c_1 -i \infty}^{c_1+ \infty}
\frac{{\rm d}s}{2\pi i}
\int_{c_2 -i \infty}^{c_2 + \infty}
\frac{{\rm d}\lambda}{2\pi i}
{\e}^{N \psi(\lambda,s)}
\label{zint}
\end{eqnarray}
where
\begin{equation}
\psi(\lambda,s) = \frac{\lambda}{\rho} + s \phi - \ln \lambda
+ \ln \left[
\int_0^\infty {\rm d}v\
 {\e}^{-sv -\lambda v^{1/p} }\right]\;.
\label{psi}
\end{equation}
The integral (\ref{zint}) may be evaluated by the saddle-point method.
The saddle-point equations coming from the conditions
$\frac{\partial \psi}{\partial \lambda}=0$
and $\frac{\partial \psi}{\partial s}=0$
read
\begin{eqnarray}
\frac{1}{\rho} &=& \frac{1}{\lambda} + \frac{
 \int_0^\infty {\rm d}v \, v^{1/p}  {\e}^{-sv -\lambda v^{1/p} }}
{ \int_0^\infty {\rm d}v\  {\e}^{-sv -\lambda v^{1/p} }}\\
\phi &=&  \frac{
 \int_0^\infty {\rm d}v \, v  {\e}^{-sv -\lambda v^{1/p} }}
{ \int_0^\infty {\rm d}v\  {\e}^{-sv -\lambda v^{1/p} }}
\end{eqnarray}
which are, of course,  precisely the grand canonical equations
(\ref{rhoscaling1}, \ref{phiscaling1}).
Thus  when the saddle point exists (i.e. in the fluid phase)
the results of the grand canonical and
micro-canonical ensembles coincide.
However, in the condensed phase $\phi > \phi_{max}(\rho)$
one can no longer  solve the saddle-point equations
for $s \ge 0$.
Therefore $Z_N(L,V)$ must be evaluated by an alternative approach.

It will be useful to consider the Laplace transform of $Z_N(L,V)$ with
respect
to the length $L$ (rather than the double Laplace transform of $Z_N(L,V)$
which generates  the grand canonical partition function).
\begin{eqnarray}
{\widetilde Z}_N(\lambda, V)
&=& \int_0^\infty {\rm d}L\ {\rm e}^{-\lambda L} Z_N(L,V)\\
&=& \prod_{i=1}^N \int_0^\infty {\rm d}v_i
\int_{v_i^{1/p} }^{\infty} {\rm d}x _i
\ {\e}^{-\lambda x_i} \delta(\sum_i v_i -V)\\
&=& \left[ \frac{\int_0^\infty {\rm d} w {\e}^{-\lambda
 w^{1/p}}}{\lambda}\right]^N
    \left[\prod_{i=1}^N \int_0^\infty {\rm d}v_i  f(v_i)\right]
   \delta(\sum_i v_i -V)
\label{Ztransf}
\end{eqnarray}
where we have defined
\begin{equation}
f(v_i) = \frac{{\e}^{-\lambda  v_i^{1/p}}}{\int_0^\infty {\rm d} w\
{\e}^{-\lambda  w^{1/p}}}.
\label{fdef}
\end{equation}
This definition ensures that the integral of $f(v_i)$ is normalised to
unity
therefore $f(v_i)$ may be considered as the probability distribution
for a positive random variable $v_i$.
Then the quantity
\begin{equation}
\Omega_N(\lambda)
 =   \left[\prod_{i=1}^N \int_0^\infty {\rm d}v_i  f(v_i)\right]
   \delta(\sum_i v_i -V)\;,
\label{Odef}
\end{equation}
manifest in (\ref{Ztransf}),
is the probability that a sum of $N$ independent positive random
variables each
distributed according
to $f$ is equal to $V$. Assuming that $f$ has finite first and second
moments
$\mu_1$ and $\mu_2$,  as it does in the  case (\ref{fdef}), where
\begin{equation}
\mu_n = \int_0^\infty {\rm d} v\ v^n f(v)\;,
\end{equation}
one can invoke some limiting
results on the sum of a large number $N$ of such random variables.
We first define
\begin{equation}
V_c = N \mu_1
\end{equation}
then the following results for sums of random variables
derived in a different context \cite{EMZ06}
will be useful
\begin{eqnarray}
\mbox{For}\quad V-V_c \sim O(N^{2/3})\qquad  \Omega_N(\lambda,V) \simeq
\frac{1}{\sqrt{2\pi N \Delta^2}}
  {\e}^{-\frac{(V-V_c)^2}{2 N \Delta^2}} \label{clt1}\\
\mbox{For}\quad V>V_c\quad  \mbox{and} \quad V-V_c \sim O(N)\qquad  
\Omega_N(\lambda,V) \simeq N f(V-V_c) \label{ldr}
\end{eqnarray}
where $\Delta^2 = \mu_2 -\mu_1^2$.
The first result is a central limit theorem
which expresses the fact that the sum is Gaussian distributed about the
mean
$V_c$.  The second  is a large
deviation result whose interpretation is that for the sum of random
variables to be equal to a value $V$,  much greater than the mean $V_c$,
one of
the random variables should be equal to $V-V_c$ to leading order
and the other $N-1$ should be of $O(\mu_1)$. In (\ref{ldr}) the factor
$f(V-V_c)$ comes from the probability of the large random variable and the
factor $N$ comes from the number of ways of choosing the large contribution
from the $N$ random variables.

Let us note that $\mu_1$ is precisely the critical
volume fraction $\phi_{max}(\rho)$ defined in the previous section.
This follows by computing the first moment of $f(v)$ in (\ref{fdef})
and comparing it to the expression of $\phi_{\max}(\rho)$ in
(\ref{eq:phimax}). Thus
\begin{equation}
V_c = N \mu_1 = N \phi_{max}(\rho)
\end{equation}
justifying the subscript $c$ (critical) for the volume $V$.

We may now obtain forms for $Z_N(L, V)$ near to criticality and in the
condensed
phase by inverting the Laplace transform
\begin{eqnarray}
Z_N(L, V)
&=& \int_{c_1-i\infty}^{c_1+i\infty} \frac{\rm d \lambda}{2 \pi i}
{\e}^{L\lambda}
 \left[ \frac{\int_0^\infty {\rm d} w\ {\e}^{-\lambda
 w^{1/p}}}{\lambda}\right]^N
\Omega_N(\lambda,V)\;.
\end{eqnarray}
Here the $\lambda$ integral may be evaluated without problem using the
saddle-point method with the saddle located (to leading order for
large $L$, large $N$ but keeping the ratio $\rho=N/L$ fixed) at
$\lambda_* = (p+1)\rho$. For $\Omega_N(\lambda^*,V)$ one can use the
results in (\ref{clt1}) and (\ref{ldr}). This gives
\begin{eqnarray}
Z_N(L,V) \simeq  \left(\frac{N}{2 \pi (p+1)}\right)^{1/2}
{\e}^{L\lambda_*} \left[ \frac{p\Gamma(p)}{(\lambda_*)^{p+1}}\right]^{N-1}
\quad\mbox{for}\quad V-V_c \sim O(N) \label{cllt2} \\
Z_N(L,V) \simeq  \left(\frac{\lambda_*^2}{2 \pi (p+1)  N}\right)^{1/2}
{\e}^{L\lambda_*} \left[ \frac{p\Gamma(p)}{(\lambda_*)^{p+1}}\right]^{N}
\frac{  {\e}^{-\frac{(V-V_c)^2}{2 N \Delta^2}} }{\sqrt{2\pi N \Delta^2}}
\label{clt2}
\quad\mbox{for}\quad V-V_c \sim O(N^{2/3})\nonumber \\
\end{eqnarray}
Therefore for $\phi \ge \phi_{max}(\rho)$ the entropy per particle,
given in the 
micro-canonical ensemble by
\begin{equation}
S = \frac{1}{N}\ln Z_N(L,V) \;,
\end{equation}
remains fixed at
\begin{equation}
S= p+1 + \ln \Gamma(p+1) - (p+1) \ln \left( (p+1)\rho \right)\;.
\end{equation}
Note that this is precisely the maximal value $S_{\max}$ in (\ref{Smax2})
computed in the fluid state. To summarize, for $p>1$, the entropy $S$
increases
monotonically with increasing $\phi$, achieves its maximal value $S_{max}$
at $\phi=\phi_{\max}(\rho)$ and then remains fixed at this value
for all $\phi>\phi_{max}(\rho)$.

We may also consider the marginal distribution $p(v)$ in the condensed
phase.
As a signature of condensation this distribution should contain a bump
at $v = V- V_{c}$.
In the micro-canonical ensemble we have
\begin{equation}
p(x,v) =  \frac{Z_{N-1}(L-x,V-v)}{Z_N(L,V)}\theta(x-v^{1/p})
\end{equation}
Integrating out $x$ yields
\begin{equation}
p(v) =  Z_N(L,V)^{-1} \int_{v^{1/p}}^L {\rm d}x\ Z_{N-1} (L-x,V-v)
\label{pvmicro}
\end{equation}
One can then substitute the asymptotic behavior of the function $Z_N(L,V)$
obtained in (\ref{cllt2}) and (\ref{clt2}) to estimate $p(v)$ in
(\ref{pvmicro}).
Omitting details, we find that
\begin{eqnarray}
\mbox{For} \quad v \ll V-V_{c}\quad
p(v) \simeq  \frac{\lambda^p_*}{ \Gamma(1+p)}  {\e}^{-\lambda_* v^{1/p}}
\label{pcrit}\\
\mbox{For} \quad v = V- V_{c} +O(N^{2/3})\quad
p(v) \simeq     \frac{1}{N} \frac{1}{\sqrt{2\pi N \Delta^2}}
  {\e}^{-\frac{(v-(V-V_c))^2}{2 N \Delta^2}}
\label{pcond}
\end{eqnarray}
The latter piece (\ref{pcond}) of $p(v)$ represents the condensate: it
has a total weight $1/N$
signifying a single condensate site and shows a Gaussian distribution
of the bump around
the excess volume $V-V_c$.


\section{The free energy functional route}
\label{sec:dft}

Having established that our problem, defined in terms of dynamical rules,
admits a factorized steady state probability with detailed balance,
we can envisage the steady state as the equilibrium state of a hard-rod
model,
where the distribution of rod lengths $\ell \equiv v^{1/p}$ is not known
{\it a priori}. This size distribution should be that which
minimizes the free energy of the system (or equivalently maximizes
the total entropy, since we only deal here with excluded volume
interactions).
Our goal in the remainder is therefore to provide a different, perhaps
more physical --but fully equivalent-- perspective, and obtain the optimal
polydispersity minimizing the free energy functional of a hard-rod system,
with the same global constraints as in section \ref{sec:model}: fixed
density $\rho=N/L$, and fixed $p$-moment of the length distribution
($\langle v \rangle = \langle \ell^p\rangle$), where $p$
is a parameter of the model, and is fixed as in previous sections. 
In doing so,
we are only interested here in the size distribution (variable $v$),
and not in the joint distribution of size and gap $x$. In this respect,
the approach of sections \ref{sec:gc} and \ref{sec:mc}
provides  more detailed information.

The problem we face is the one-dimensional analogue of the optimal packing
of polydisperse hard sphere fluids addressed in Refs.
\cite{ZBTCF99,B00,BC01},
which exhibits an unexpected condensation transition. The free energy
functional
approach put forward here is akin to that used in \cite{ZBTCF99}, with
nevertheless the interesting feature that in the present 1D geometry,
exact results can be obtained. On the other hand, approximations of
Percus-Yevick type or more involved treatments
were necessary in Refs \cite{ZBTCF99,B00,BC01}.
Indeed, the problem boils down to finding the size dependence of the
chemical potential of a given species in a polydisperse mixture,
which is not known exactly for hard discs or hard spheres.

\subsection{Free energy functional for polydisperse hard rods}

We work in the canonical ensemble where a distribution of $N$ hard
rods at temperature $T$ occupies an available length $L$.
Their total length is denoted $L_p$ and defines the line coverage
(or packing fraction) $\eta  = L_p/L = \rho \langle \ell \rangle$ where
$\rho = N/L$ is the density and $\langle \ell \rangle$ is the mean size.
To establish the connection with the $p$-spheres discussed earlier,
we can view each rod of size $\ell$ as a (hyper)sphere
constrained to move on a line, and having ``volume''
$v=\ell^p$.

Writing the ideal entropy of a multicomponent and discrete system
is straightforward \cite{SS82}, but the limit of a continuous distribution
requires some care (see the discussion below). For hard rods,
the free energy functional may be written as a sum of ideal and excess
contributions \cite{TheOtherEvans}:
\begin{equation}
\beta{\cal F}\{W\} = N\, \int {\rm d}\ell \, W(\ell) \left[ \ln \left(\Lambda^2
\rho W(\ell) j_\ell(\ell)
\right) -1 \right] + \beta{\cal F}_{\hbox{\scriptsize ex}}\{W\}.
\label{eq:functional}
\end{equation}
where $\Lambda$ is an irrelevant length scale, $\beta$ is the inverse
temperature, and $W$ is the length probability distribution function
(such that $\int d\ell\ \ell W(\ell) = \langle \ell \rangle$).  
The excess free energy for a homogeneous system of 
hard rods is known exactly, 
$ \beta{\cal F}_{\hbox{\scriptsize ex}}\{W\}
=- \ln(1-\eta)$~ \cite{SS82}, 
and can be generalized to inhomogeneous situations \cite{TheOtherEvans}.
In Eq.(~\ref{eq:functional}), 
the function $j_\ell(\ell)$ --yet to be specified--
ensures that different choices for the labelling of the particles
will lead to the same optimal length distribution $W(\ell)$.
One could indeed choose $v=\ell^p$ (or say any other power)
as a working variable, with an associated ``labelling'' function
$j_v(v)$ and a probability distribution function
$W_v$ such that $W(\ell) = p \,\ell^{p-1} W_v(\ell^p)$.
Enforcing the consistency of both descriptions imposes
$j_\ell(\ell) =  \ell^{p-1} j_v(\ell^p)$ (up to an irrelevant
prefactor), where the factor $\ell^{p-1}$ is the Jacobian of the
transformation
$\ell \to \ell^p$. The natural labelling  for the particles,
i.e. that which gives a constant function $j$, follows from
the way polydispersity is sampled (see e.g. \cite{ZBTCF99,B00}),
and reflects the dynamics of the system.
In the following Monte Carlo simulations, we will attempt to change the
size of two particles $\ell_1$ and $\ell_2$
selected at random by adding a small increment to $\ell_1^{p}$
and conversely subtracting the same quantity from  $\ell_2^{p}$,
in accordance with the rule specified in section \ref{sec:model}
and more precisely with the requirement that $b(v)=1$.
This corresponds to a flat function $j_v$, and hence to
$j_\ell = \ell^{p-1}$. 
However, given an arbitrary $b(v)$ (not constant) in the microscopic rate
in Eq. (\ref{w}), it is far from evident how
to choose an appropriate labelling function $j_v(v)$ in the
macroscopic description in Eq. (\ref{eq:functional}), 
that would precisely correspond to this
microscopic rate.

Minimization
of ${\cal F}$---with the two constraints of normalization
$\int {\rm d} \ell  W(\ell)=1$ 
and fixed $p$-moment $\int {\rm d} \ell  \ell^p W(\ell)$, that can be
included by two Lagrange multipliers---leads to the functional form
of the optimal size distribution $W^*$, which should then
coincide with the probability distribution function
 of $v^{1/p}$, where the statistics
of $v$ is given by Eq. (\ref{pv}).
Since the explicit expression of $\cal F$ is known,
see above,
this provides a first angle of attack to our problem.
However, interesting information
follows from more microscopic considerations and scaling
arguments, as becomes clear below.
Note that the constraint that the line coverage should not
exceed unity is not explicitly considered, but is implicitly
encoded in the excess contribution to the free energy functional
(\ref{eq:functional}).

\subsection{Optimal polydispersity distribution}
\label{sec:opd}
We take advantage of the fact that any
infinitesimal change in $W^*$ has a vanishing free
energy cost $\delta {\cal F}$, provided the
two aforementioned constraints are satisfied.
We proceed in two steps: {\em a)} a given rod of arbitrary length
$\ell_0$ is expanded : $\ell_0 \to \ell_0 + \delta \ell_0$; {\em b)}
all particles are rescaled ($\ell \to \alpha\, \ell$) in order to
fulfil the constraint of a conserved $p$-th moment.
In the first step, this moment changes by an amount
\be
\frac{1}{N} \,\left[(\ell_0+\delta \ell_0)^p - \ell_0^p \right] \,
\simeq \,
\frac{1}{N} \, p\, \ell_0^{p-1}\, \delta \ell_0
\ee
while in the second stage, it changes by
$(\alpha^p-1) \langle \ell^p \rangle \simeq p (\alpha-1) \langle \ell^p
\rangle$.
Enforcing the conservation of $\langle \ell^p \rangle$ then imposes
\be
\alpha \,\simeq \, 1 -
\frac{\ell_0^{p-1} \delta \ell_0}{N \langle \ell^p \rangle}
\label{eq:alpha}
\ee

In stages {\em a)}+{\em b)}, the distribution function changes by
an amount
\be
\delta W =  \frac{1}{N}\,\left[\, \delta \left( \ell-\ell_0-\delta
\ell_0\right)
\,-\, \delta\left(\ell-\ell_0\right) \, \right] + (1-\alpha)
\, \frac{{\rm d} [\ell W]}{{\rm d}\ell}.
\ee
It proves convenient to express the variation of the ideal part of
the free energy in terms of the probability distribution function $W_v$ of
$v$. This yields, with $v_0=\ell_0^p$
\be
\delta {\cal F}_{\hbox{\scriptsize id}} =  k_B T \left\{
\frac{W_v'(v_0)}{W_v(v_0)} +
\frac{1}{\langle v\rangle}\right\}
\delta v_0.
\label{eq:dF}
\ee
The excess contribution variation --which is most conveniently
expressed in terms of $\ell$ rather than $v$--
 is in addition exactly the
reversible work required to perform the changes under consideration.
In step {\em a)}, we note that this work is the same as if a confining
``wall'' would be displaced a distance $\delta \ell_0$ thereby compressing
the system. We therefore have
\be
\delta {\cal F}_a \,=\, P \,\delta \ell_0,
\ee
where $P$ is the pressure of the system.
This relation is specific to the one dimensional case, and at the
root of important simplifications as compared to higher dimensions
\cite{TP08}.
On the other hand, it is a general result that in any space
dimension, the work performed during step {\em b)} is
(see e.g. appendix B of \cite{ZBTCF99})
\be
\delta {\cal F}_b \,=\, -\frac{P_{\hbox{\scriptsize ex}}}{\eta} \delta L_p
\ee
where $P_{\hbox{\scriptsize ex}} = P-\rho kT$ is the excess pressure and
$\delta L_p$ is the total volume change of the particle upon
the rescaling $\ell \to \alpha\, \ell$:
$\delta L_p = (\alpha-1) L_p = (\alpha-1) N \langle \ell \rangle$.
Making use of Eq. (\ref{eq:alpha}), we finally have :
\be
\frac{\delta {\cal F}_{\hbox{\scriptsize ex}}}{\delta \ell_0} =
P  -\frac{P_{\hbox{\scriptsize ex}}}{\eta} \,
\frac{\ell_0^{p-1}\langle \ell \rangle}{\langle \ell^p \rangle}.
\ee
Combining this with Eq. (\ref{eq:dF}) supplemented by the
requirement that $\delta {\cal F}=0$ for $W=W^*$
(or equivalently $W_v=W_v^*$) gives
\be
\frac{ {W_v^*}'(v_0) }{W_v^*(v_0)} + \frac{1}{\langle v\rangle} +
\frac{P}{p} \,v_0^{1/p-1} - \frac{P_{\hbox{\scriptsize ex}}}{\eta} \,
\frac{\langle \ell \rangle}{p\,\langle \ell^p \rangle} = 0.
\ee
After integration with respect to $v_0$, we arrive at
\be
W_v^*(v) \, \propto \, \exp\left\{
-\beta P v^{1/p} \,+ \,
\left(\frac{\beta P_{\hbox{\scriptsize ex}}\langle \ell
\rangle}{p\,\eta} -1\right)\,
\frac{v}{\langle \ell^p \rangle}
\right\}.
\label{eq:px}
\ee
This is, expectedly, the very same form as obtained in section
\ref{sec:gc}, see Eq. (\ref{pv}). In addition, Eq. (\ref{eq:px}) above
makes explicit the connection between the chemical potentials
$\lambda$ and $s$ appearing in Eq. (\ref{pv}) and intensive
thermodynamic  quantities,
for example $\lambda = \beta P$, see Eq. (\ref{etadef1}).
Equivalently, in terms of the $\ell$ variable, we obtain
\be
W^*(\ell) \,= \,{\cal A} \, \ell^{p-1} \, \exp\left\{
-\beta P \ell \,+ \,
\left(\frac{\beta P_{\hbox{\scriptsize ex}}\langle \ell
\rangle}{p\,\eta} -1\right)\,
\frac{\ell^{p}}{\langle \ell^p \rangle}
\right\}
\label{eq:pdfW}
\ee
where ${\cal A}$ is a normalization factor.
As alluded to earlier, it is noteworthy here
that the pressure is exactly related to the density $\rho$ through
\cite{Tonks,SS82,TP08}
\be
\beta P \, = \, \frac{\rho}{1-\eta} \qquad \hbox{where}
\quad \eta = \rho \,\langle \ell \rangle.
\label{eq:press}
\ee

One can note that the low density behaviour of $W^*$ allows for some
consistency test of our prediction. When the density vanishes,
one has $P\to 0$ and $\beta P_{\hbox{\scriptsize ex}}/\eta \to 0$,
so that
\be
W^*(\ell) \, \propto \,\ell^{p-1} \exp\left(-\frac{\ell^{p}}{\langle
\ell^p \rangle}
\right),
\label{eq:pdflown}
\ee
hence an exponential distribution of $v=\ell^p$ (the conserved quantity).
Such an expression also immediately follows from direct minimization
of the functional (\ref{eq:functional}), restricted to its ideal
contribution
(${\cal F}_{\hbox{\scriptsize ex}}=0$).
We recover here the expression obtained in section \ref{sec:gc}
in the large $p$
limit. As mentioned earlier, we also recover
the results reported in an econophysics context \cite{Dragu}
for a simple model where agents
act as our ideal particles, exchanging random
amounts of a quantity $v$ (money) in binary encounters.

With a weight function such as (\ref{eq:pdflown}), one immediately finds
\be
\frac{\int_0^\infty \ell^p \,W^*(\ell)\, {\rm d}\ell}{\int_0^\infty
W^*(\ell)\,  {\rm d}\ell}
= \langle \ell^p \rangle,
\label{eq:req}
\ee
as it should (the quantity $\langle \ell^p\rangle$ appearing
on the l.h.s. of (\ref{eq:pdflown}) is therefore indeed the moment of
order $p$ of the distribution).
This shows the consistency of our distribution function
in the low density limit.
The corresponding mean size follows, assuming again the
low density form (\ref{eq:pdflown}) :
\be
\frac{\int_0^\infty \ell \,W^*(\ell)\,  {\rm d}\ell}{\int_0^\infty W^*(\ell)\,
 {\rm d}\ell}
= \Gamma(1+1/p) \,\,\langle \ell^p \rangle^{1/p},
\label{eq:meanllown}
\ee
With the standard ``functional route'' alluded to earlier
(i.e. direct minimization of Eq.(\ref{eq:functional})) which
does not provide explicitly Lagrange multipliers, the requirement
(\ref{eq:req}) together with normalization
would determine those multipliers for any density.

In the remainder, the star superscript will be omitted to refer to
the optimal distribution $W^*$, without ambiguity.

\subsection{The condensation transition}\label{sec:condrev}

We revisit here the condensation transition brought to the fore in section
3, in the more liquid-state language of hard rods.
Since the excess pressure following from (\ref{eq:press}) fulfils the
identity
$P_{\hbox{\scriptsize ex}}/\eta=P$, as can be readily checked,
the distribution (\ref{eq:pdfW}) can be rewritten more explicitly as
\be
W(\ell) \,= \,{\cal A} \, \ell^{p-1}\,\exp\left[
-\frac{\rho \ell}{1-\eta} \,+ \,
\left(\frac{\eta-\eta_c}{\eta_c(1-\eta)}
\right)\,
\frac{\ell^{p}}{\langle \ell^p \rangle}
\right],
\label{eq:W2}
\ee
with $\eta_c = p/(1+p)$.
For $p\leq 1$, this distribution is normalizable for all line coverages  $\eta$ (given that $\eta<1$).
This is no longer the case for $p>1$, provided that $\eta>\eta_c$,
where $\eta_c$ will be referred to as the critical line coverage.
For $p>1$ and $\eta>\eta_c$, the divergent behaviour of
distribution (\ref{eq:W2}) at large $\ell$ is indicative of the
formation of a macroscopic
aggregate with size $L_0$. The scenario is identical to that inferred
from the observations
of Ref. \cite{ZBTCF99} and worked out in \cite{BC01}. The system
relaxes by transferring ``volume'' to the aggregate, which
effectively acts as a piston and coexists with a
polydisperse mixture ${\cal M}$ confined in a region of size $L-L_0$.
In this region, the size distribution obeys Eq. (\ref{eq:W2}),
where $\eta$ is no longer the total line coverage fraction, but should be
replaced
by the line coverage  $\eta_{_{\cal M}}$ in the region free of aggregate.
The different line coverages in the problem are connected through
\be
\eta_{_{\cal M}} = \frac{\eta-\eta_0}{1-\eta_0},
\ee
where $\eta_0=L_0/L$ is the line coverage of the aggregate.
We show in Appendix \ref{app:B} that
\be
\eta_0 = \left\{
\begin{array}{c l}
 0 &\hbox{ for } \eta<\eta_c\\
\displaystyle\frac{\eta-\eta_c}{1-\eta_c} & \hbox{ for } \eta>\eta_c.
\end{array}
\right.
\label{eq:op}
\ee
The line coverage  $\eta_0$ may be considered as an order parameter
for the transition. The key ingredient to arrive at
Eq. (\ref{eq:op}) is that the fluid outside the condensate
(assuming the latter species forms) is critical, in the available
length $L-L_0$. It therefore has line coverage $\eta_{\cal M} =
\eta_c$, and size distribution
\be
W(\ell) \, \propto \, \ell^{p-1}\,\exp\left[
-\frac{\rho \ell}{1-\eta_c}
\right].
\label{eq:pdfcrit}
\ee
We note finally that the case $p\to \infty$ is specific in that
we then have $\eta_c\to 1$ (while $\phi^*_c\to 4/e$). Since the
hard core constraint imposes $\eta<1$, we see here why the large
$p$ limit reduces in fact to the ideal gas case, where $\ell^p$
is exponentially distributed and no transition occurs.


\section{Comparison with simulation results}
\label{sec:sim}

In order to verify analytical predictions it is useful to compare with
numerical results. For example,  numerical studies of condensation have in
the past revealed important information  about when the asymptotic behaviour
predicted analytically actually emerges in a  finite system.

\subsection{Control parameter and critical point}
To analyze in more detail the scenario at work and put our predictions
to the test, we perform Monte Carlo simulations.
We first have to introduce
a relevant control parameter (we emphasize that except when $p=1$,
the  line coverage or packing fraction $\eta$ (\ref{etadef})
is not a conserved quantity, but is self-consistently determined).
Since both the density and $\langle\ell^p  \rangle$ are
conserved variables, we will use the dimensionless reduced volume fraction
\be
\phi^* \,=\, \rho \langle\ell^p  \rangle^{1/p},
\ee
introduced in eq.~(\ref{phi*def}).
A simple convexity argument shows that for $p>1$,
$\phi^*>\eta$, while the reverse holds for $p<1$.
Having chosen $\langle\ell^p  \rangle^{1/p}$ as our relevant length scale,
we also introduce a reduced pressure
\be
P^* \, \equiv \, \beta P \langle\ell^p  \rangle^{1/p} =
\frac{\phi^*}{1-\eta}
= \frac{\phi^*}{1-\rho \langle\ell  \rangle},
\label{eq:eofs}
\ee
and a rescaled length
\begin{equation}
\ltil = \ell / \langle \ell^p\rangle^{1/p}\;
\end{equation}
which, from (\ref{eq:W2}), has size distribution
\be
W(\ltil) \,= \,{\cal A'} \, \ltil^{p-1}\,\exp\left[
-\frac{\eta}{1-\eta} \frac{\ltil}{\langle \ltil\rangle} \,+ \,
\left(\frac{\eta-\eta_c}{\eta_c(1-\eta)}
\right)\,
(\ltil)^{p}
\right].
\label{eq:Wtilde}
\ee

It is in general not possible to relate explicitly the reduced density
$\phi^*$ to the line coverage $\eta$, except
at the critical point $\eta=\eta_c$, since then $W$ takes
a pure exponential shape (up to the algebraic prefactor).
From (\ref{eq:W2}), we have there
\be
\langle\ell^k  \rangle \,\stackrel{\eta=\eta_c}{=}\,
\frac{\Gamma(p+k)}{(\beta P)^k\, \Gamma(p)}
\quad \hbox{for} \quad k \geq 0
\ee
so that
\be
P^* \,\stackrel{\eta=\eta_c}{=}\, \left(
\frac{(2p-1)!}{(p-1)!}
\right)^{1/p} = (1+p) \phi^*.
\ee
Starting from $\phi^*=0$
when $\eta=0$, the reduced density $\phi^*$  increases with $\eta$ and
reaches the value
\be
\phi^*_c \,=\, \frac{1}{1+p} \,\left(\frac{(2p-1)!}{(p-1)!}
\right)^{1/p}
\ee
when $\eta=\eta_c$. 
This corresponds exactly to  the threshold obtained in section
\ref{sec:mc}, see Eq. (\ref{eq:phimax}).
For $p>1$, $\eta_c$ signals the onset of the formation of the condensate.
For $p<1$, $\eta>\eta_c$ only signals a region where the sub-dominant
term in $\ell^p$
in the exponential (\ref{eq:W2}) changes sign. The size distribution may
therefore
change from the unimodal shape found at low line coverage to a
bimodal form
when $\eta$ exceeds some threshold, itself larger than $\eta_c$. On the
other hand,
the onset of bimodality for the distribution $W_v$ of $v=\ell^p$ is
$\eta_c$.

When no condensate forms (i.e. $p \leq 1$ or $\phi<\phi_c$ if $p>1$),
we compute numerically the probability distribution function of $\ltil$
as follows.
For a given value of the reduced density $\phi^*$, the mean value
$\langle \ltil\rangle$ appearing on the rhs in (\ref{eq:Wtilde}) is
determined
self consistently by enforcing that it should coincide with the first
moment of the distribution
having statistical weight (\ref{eq:Wtilde}). Note that
$\phi^*=\eta/\langle \ltil\rangle$.
Alternatively, we may also
impose the second self-consistency requirement that $\langle \ltil^p
\rangle=1$.
We have systematically checked that both routes provide the same result
for \smash{$\langle \ltil\rangle$}, from which the different quantities of
interest may be computed. This illustrates the consistency of the
functional form
(\ref{eq:Wtilde}). With such a procedure, the solution found numerically
is always unique; it will be compared against the results of Monte Carlo
simulations in the next section.

In situations where a condensate is expected ($\phi>\phi_c$ and $p>1$),
an explicit prediction between the  control  parameter $\phi^*$
and the condensate size (or more precisely
condensate line coverage) can be derived
from the remark that the fluid phase outside the condensate is critical.
In other words,
\be
\phi^*_c = \frac{N-1}{L-L_0} \left(\frac{1}{N} \sum_{i \in {\rm fluid}} \ell_i^p
\right)^{1/p}.
\ee
On the other hand, the global reduced density (including, thus, the
condensate) reads
\be
\phi^* = \frac{N^{1-1/p}}{L} \left(L_0^p +\sum_{i \in  {\rm fluid}} \ell_i^p
\right)^{1/p}.
\ee
The two above equations allow us to compute $\eta_0$, the condensate
line coverage, through
\be
\phi^{*\,p} \,=\, N^{p-1} \,\eta_0^p + \left[\phi^*_c \,(1-\eta_0)
\right]^p.
\label{eq:eta0phi}
\ee
Hence, for large $N$, $\eta_0 \to 0$, and we have
\be
\eta_0^p \,\sim \, \frac{\phi^{*\,p}-\phi^{*\,p}_c}{N^{p-1}}.
\label{eq:eta0pred}
\ee

\subsection{Monte Carlo simulations}

To test our predictions, we have implemented Monte Carlo simulations,
closely following the algorithm used in Refs \cite{ZBTCF99}.
$N$ hard rods with different sizes are confined on a line of length $L$;
a first type of move amounts to randomly selecting a particle,
and randomly translating it.  A second kind of move allows the system
to relax its size distribution, and sample polydispersity.
Two particles are selected at random in the system; the size $\ell_1$ of
particle 1 is expanded at the expense of the size of particle 2,
so that $\ell_1^p + \ell_2^p$  is constant:
$\ell_1^p \to \ell_1^p + \Delta ; \ell_2^p \to \ell_2^p - \Delta,
$
where the increment $\Delta$ is drawn from a distribution $w$ such that the
typical value is small compared to $\langle \ell^p \rangle$.
Both types of moves are accepted provided they do not lead to
any overlap between the rods and for the second kind, provided
the shrunk rod does not have a negative length. These rules correspond 
to the model defined in section \ref{sec:model}, in particular to the case where
the volume exchange rate  $w(\Delta,v)$ of section 2 does not depend on $v$.

\begin{figure}
\includegraphics[width=0.46\textwidth,clip=true]{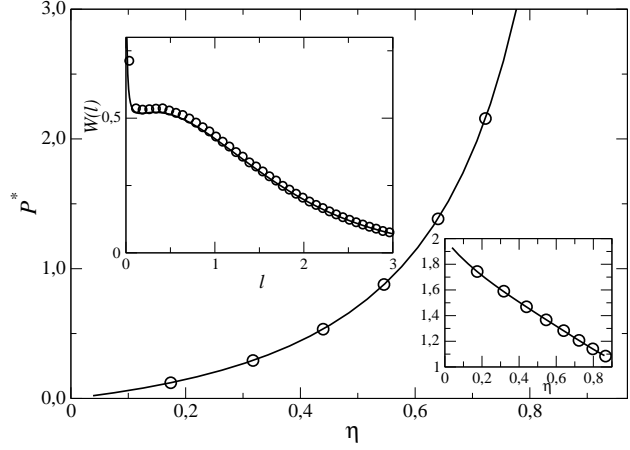}
\caption{\label{fig:p0.5} Equation of state as a function of line coverage
$\eta$ in the case where the exponent defining the conserved moment is
$p=1/2$.
The symbols are for the Monte Carlo results, and the curve is for
Eq. (\ref{eq:eofs}).
A characteristic length distribution is shown in the upper inset
for $\eta \simeq 0.72$ (corresponding to $\phi^*=0.6$).
The distribution (\ref{eq:Wtilde}) is compared to its Monte Carlo
counterpart.
The lower inset shows the first moment
$\langle \widetilde \ell\rangle$ as a function of line coverage.
Here,
$\eta_c=1/3$ for which
$\phi^*_c = 2/(3 \pi) \simeq 0.212$ and $\langle \ltil\rangle_c=\pi/2$.
Monte Carlo data with $N=1000$ particles are shown with the circles.   }
\end{figure}

\begin{figure}
  \includegraphics[width=0.46\textwidth,clip=true]{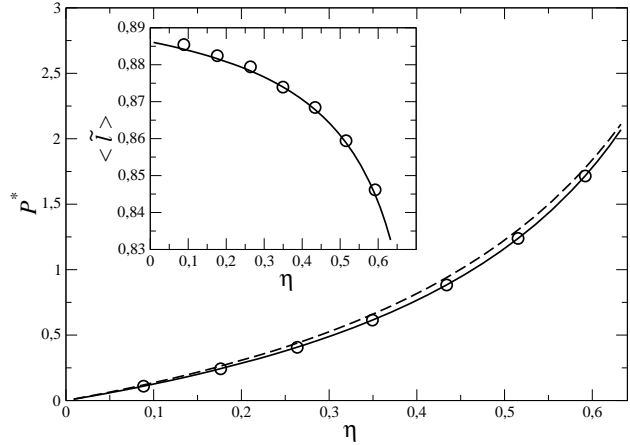}
  \caption{\label{fig:p2} Equation of state as a function of 
line coverage 
  $\eta$ when $p=2$, in which case the critical line coverage
  $\eta_c=2/3$ corresponds to
  $\phi^*_c=\sqrt{2/3}\simeq 0.816$.
  The exact pressure is shown by the continuous line, while the dotted
  curve is for the crude approximation where $\langle \widetilde
\ell\rangle$
  is assumed density independent and equal to its critical value $\sqrt{
2/3}$,
  which leads to $P^*=\sqrt{3/2}\,\eta/(1-\eta)$. Circles : Monte Carlo
results
  with $N=1000$.
  The inset shows the
  density dependence of $\langle \widetilde \ell\rangle$.
   }
\end{figure}

Typical results are shown in Figs. \ref{fig:p0.5} and \ref{fig:p2},
that are both for cases without condensation.
In all figures, the data gathered from the Monte Carlo simulations are
shown with
the symbols, while the predictions are shown by the continuous curves.
The agreement theory/simulations is very good (see in particular the
distribution function in the upper inset of Fig. \ref{fig:p0.5}).
From Eq. (\ref{eq:meanllown}), we have that in the low density limit,
$\langle\widetilde \ell\rangle =2$ for $p=1/2$ (see the lower inset
of Figure \ref{fig:p0.5}), and
$\langle\widetilde \ell\rangle = \sqrt{\pi}/2 \simeq 0.886$ for
$p=2$ (which can indeed be seen in the inset of Figure \ref{fig:p2}).

\begin{figure}
  \includegraphics[width=0.46\textwidth,clip=true]{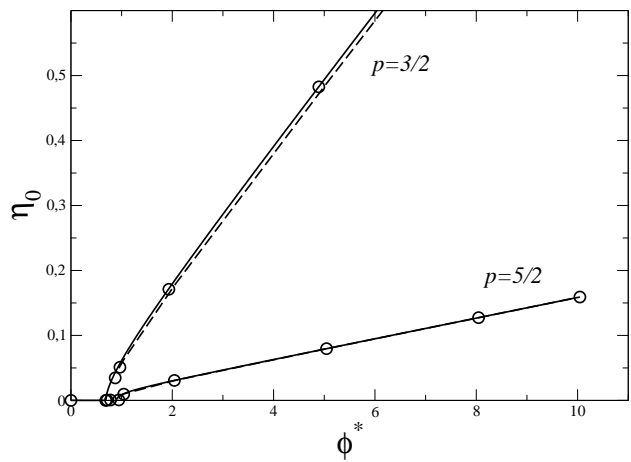}
  \null\vskip 1ex
  \caption{\label{fig:condensate}
  Packing fraction (line coverage)
  of the condensate as a function of  reduced volume fractions
  $\phi^*$, for $p=3/2$ (lower sets) and $p=5/2$ (upper sets).
  The symbols have been obtained by computing the mean
  size of the largest particle observed in Monte Carlo simulations,
  and the curve shows Eq. (\ref{eq:eta0phi}). For the present parameters
  ($N=1000$), the approximation provided by Eq. (\ref{eq:eta0pred}) proves
  quite accurate (dashed line, hardly distinguishable from the
  continuous curve for $p=5/2$).
   }
\end{figure}

We now turn to the cases where a condensate should form.
It may then be difficult, from a practical point of view, to
distinguish such a big particle from others belonging to the
tail of the size distribution.
Another difficulty comes from the fact that the condensate line fraction
$\eta_0$ may be small for large $N$.  
Attention must be paid to the size of the condensate that is expected.
Eq. (\ref{eq:eta0pred})  
indicates that the condensate line coverage
scales with $N$ like $N^{-1+1/p}$; however, the typical rod size
in the fluid phase concomitantly exhibits a faster decay in $1/N$, so
that there is always a clear separation condensate/fluid.
In our simulations, we have followed for $p>1$
the biggest particle (size $L_0$) in the simulation box (line). As can
be seen
in Fig. \ref{fig:condensate}, the corresponding line fraction
$\eta_0=L_0/L$ closely follows our prediction (\ref{eq:eta0phi}).
We therefore conclude that Fig. \ref{fig:condensate} proves the
existence of the condensate for $\phi^*>\phi^*_c$.

\begin{figure}
  \includegraphics[width=0.46\textwidth,clip=true]{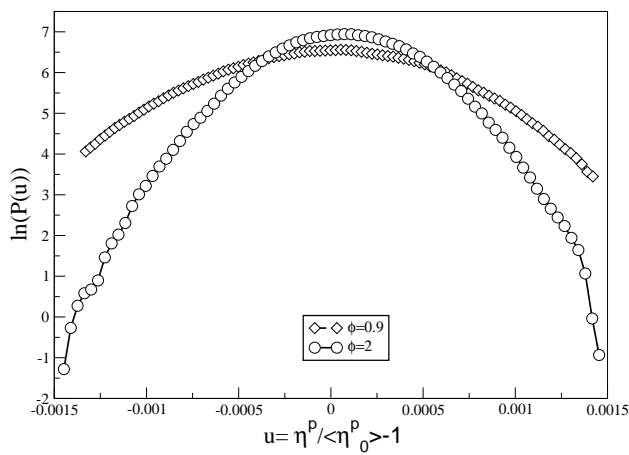}
  \caption{\label{fig:aggrdist} Linear-log plot of the
  aggregate size distribution for $p=3/2$ and two reduced volume fractions
  ($\phi^*=0.9$ and 2). The data are obtained following over time
  the different sizes taken by the (unique) condensate. Here, $P$ denotes the
  probability distribution function of $y=\eta^p/\langle \eta_0^p\rangle -1$.
   }
\end{figure}

\begin{figure}
  \includegraphics[width=0.46\textwidth,clip=true]{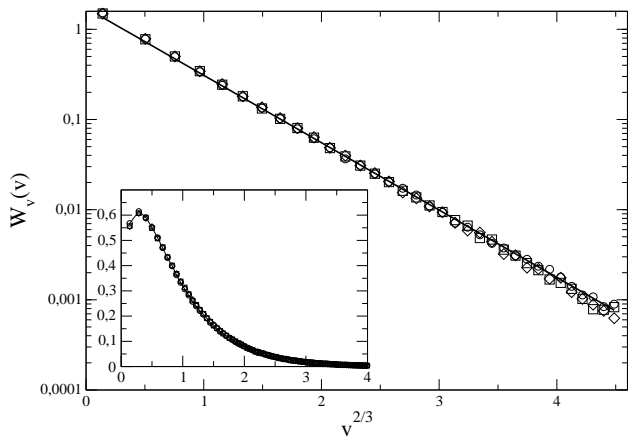}
  \caption{\label{fig:p1.5} Linear-log plot of the
  probability distribution function $W_v$ of
  particle ``volumes'' $v=\ltil^p$ as a function of $\ltil$,
  for $p=3/2$ and several reduced
  volume fractions  beyond the condensation transition (circles for
  $\phi^*=2$, squares for $\phi^* =5$ and diamonds for $\phi^*=8$).
  Here $\phi^*_c=4 (2/\pi)^{1/3}/5 \simeq 0.688$. The continuous
  curve shows the prediction of Eq. (\ref{eq:pdfcrit}).
  The inset shows the probability
  distribution of the variable $\ltil$ instead of $\ltil^p$,
  again as a function of $\ltil=v^{1/p}$.)
   }
\end{figure}

To confirm the results of section 4 where the condensate 
volume distribution is calculated, see Eq.  (\ref{pcond}), we measured
the fluctuations in the condensate volume.
We find that that the distribution $p(v)$ turns out to be  Gaussian, as predicted
in section 4 (see Fig. \ref{fig:aggrdist}).
Another prediction is that the
fluid phase outside the condensate should have a size distribution
of the form (\ref{eq:pdfcrit}), irrespective of the reduced volume fraction
provided $\phi^*> \phi^*_c$. This expectation is fully consistent
with the Monte Carlo results, see Fig. \ref{fig:p1.5} for $p=3/2$.
For convenience,
we have considered in the main graph the distribution of
$\ltil^p$, since it is predicted
to be a pure exponential. We have performed a similar analysis
at $p=5/2$, for various volume fractions beyond the critical one,
and the same conclusion holds with again a critical fluid phase,
and a size distribution (excluding the condensate)
that does not depend on the imposed
volume fraction $\phi$. The fluid phase can only accommodate a well defined
finite fraction of the total volume (or length), so that when
$\phi^*$ increases above $\phi^*_c$, the extra volume is transferred
to the aggregate (see Fig. \ref{fig:condensate}).

\vskip 2ex

\section{Conclusion}
\label{sec:conc}

We have considered a simple stochastic model of mass transport where
a real-space condensation takes place upon increasing the 
volume fraction.
The system studied consists of $p$-spheres, constrained to move on a ring,
with hard core constraints : the left to left particle
distance $x_i$ between $p$-spheres $i$ and $i+1$ should exceed the diameter
$\ell_i$ of $p$-sphere $i$ (see Fig. \ref{fig1}).
The latter quantity is itself related to the volume of the sphere
through $\ell_i=v_i^{1/p}$, where $p$ has been retained as a parameter.
The total volume of the particles, $\sum_i v_i$, is fixed, so that
the present model involves two conserved quantities
($\sum_i x_i$, the total length available, and the total volume
$V=\sum_i v_i $).
The dynamics comprises diffusion (which can be seen as a stochastic
exchange
of the quantity $x$ between neighbouring particles), and a stochastic
exchange
of volume. When $p>1$ and beyond a critical density that has been
worked out explicitly, a particle of macroscopically large ``mass''
(large volume) appears,
carrying a finite fraction of the system total volume $V$, surrounded by
a critical fluid phase, with a size distribution independent of the total
density.

Under mild conditions pertaining to the sampling of the conserved
quantities,
we have shown that the system admits a factorized steady state probability
density, with detailed balance between allowed configurations
(essentially those with no overlaps between the $p$-spheres). This allows us
to use an alternative description of the system in terms of
finding the equilibrium distribution of an ensemble of hard
rods, that do not have a quenched size distribution, but can exchange
length
provided the global constraint $\sum_i \ell_i^p$ remains fixed.
Formulated as
such, the problem appears as the one dimensional analogue of the optimal
polydispersity studies of hard discs and hard sphere fluids
\cite{ZBTCF99,B00,BC01,CS02} (for instance, the problem
studied in \cite{ZBTCF99} corresponds to the three dimensional hard sphere
situation with $p=1$). The phase transition reported in
\cite{ZBTCF99,B00,BC01,CS02}
can therefore be seen as a condensation arising from constraints in the
configuration space.

We have characterized the scenario at work for this
transition, and obtained analytically the probability distributions
of volumes, gap distance $g$, and left to left particle distance $x$.
Restricting to the volume distribution, we have shown that a density
functional approach supplemented with scaling considerations allows to
recover the results derived from the stochastic processes viewpoint
(working out the consequences of the factorization property of the steady
state). In this respect, the problem is to find the polydispersity
of a hard rod system that will minimize its free energy, given that the
total number of rods is fixed as that the $p$ moment of the diameter
distribution is fixed as well. The consistent predictions of both
approaches
have finally been successfully tested against Monte Carlo simulations.
It is remarkable that this optimality problem leads to a phase
transition in a 1D system, where the condensate size can be considered
as an order parameter
(note that the constraints imposed introduce
a global coupling between all particles). It should also be emphasized that
whereas the higher dimensional systems in two or three dimensions could
exhibit
a transition for $p=1$ (with the convention that the conserved
quantity for hard spheres in $d$ dimensions is
$\langle\ell^{dp}\rangle$), such a value for $p$-spheres on a line
turns out to be critical with however no transition observed:
a macroscopic aggregate can only form provided $p>1$.

An important question that remains is that of the dynamic pathways
to condensation, that is,
how does a single condensate emerge from some given initial condition
for the $p$-sphere volumes.
As the condensation is driven by constraints in configuration space 
the dynamics too might  be strongly affected by constraints, possibly
producing, for example, entropic barriers. This issue remains to be explored.

Acknowledgements~: M.R.E. and I. P. thank
CNRS and LPTMS, Orsay  for hospitality.
E.T. acknowledges the hospitality of the Fundamental 
Physics Department of the University of Barcelona,
where part of this work was performed. I.P. acknowledges 
financial support from MICINN (FIS2008-04386) and DURSI (2005SGR00236).
We would also like to thank M.I. Garc\'ia de Soria and P. Maynar
for useful discussions.


\begin{appendix}
\section{}
\label{app:A}
In this appendix we show that the grand canonical entropy $S(u)$
defined in (\ref{Su}) as
\begin{equation}
S(u) = -(p+1)\,\ln \rho -(p+1)\,
\ln F_{\rho}(u)+\ln H(u) +F_{\rho}(u)+u\,F_{\phi}(u).
\end{equation}
is maximised when the scaling variable $u$ takes the values zero.

Expanding about $u=0$ to  second order in $u=0$ and using the definitions
of $H(u)$, $F_\rho(u)$ and $F_\phi(u)$ given in section 3.2 we obtain
\begin{equation}
S(u) = \left(1 + p - (1 + p) \ln(1 + p) + \ln \left[ \Gamma(1 + p)\right]\right)
+ a_1 u + a_2 u^2 +O(u^3)
\end{equation}
where 
\begin{eqnarray}
a_1&=&0\\ 
a_2 &=& [(1+p+p^2)\Gamma^2(2p)-(1+p)\Gamma(p)\Gamma(3p)]/[2(1+p)\Gamma^2(p)]
\end{eqnarray}
The first equation  shows that $u=0$ is a stationary point.
The function
\begin{equation}
 c(p)= (1+p+p^2)\Gamma^2(2p)-(1+p)\Gamma(p) \Gamma(3p)
\end{equation}
 is maximised
at $p =0.614413$  where its value $c(p)=-0.566304$ is negative.  Hence $c(p)$
is negative for all $p>0$, which implies $a_2<0$ for all $p>0$
proving indeed that at $u=0$, $S(u)$
is a maximum.

\section{}
\label{app:B}
To find out what does determine $\eta_0$ (and thus $\eta_{_{\cal M}}$)
for a given value
of $\eta$, we reconsider the free energy functional
(\ref{eq:functional}), supplemented
with two Lagrange terms to fulfil the constraints
\cite{ZBTCF99,B00,BC01}
\be
{\cal R}(W)=
\beta {\cal F}(W) \,+\, N {\cal L}_0 \int W(\ell)  {\rm d}\ell \,+\, N {\cal
L}_p \, \left(
\int \ell^p W(l) \, {\rm d}\ell+ L_0^p
\right).
\ee
We have to minimize this expression with respect to $W$ and $V_0$.
Stationarity with respect to $W$ leads to an expression of the form
(\ref{eq:W2}):
\be
\log [\ell^{1-p} W(\ell)] + \frac{\delta \beta {\cal
F}_{\hbox{\scriptsize ex}}}{N\delta W(\ell)} + {\cal L}_0
+{\cal L}_p \ell^p =0,
\label{eq:statW}
\ee
while the derivative with respect to $V_0$ reads
\be
\frac{\partial {\cal R}}{\partial L_0} \,=\,
\frac{\partial \beta {\cal F}}{\partial L_0} + p N {\cal L}_p
L_0^{p-1}.\index{}
\label{eq:statL0}
\ee
The free energy of the system
is a function of $L-L_0$, since a constituted aggregate does not
contribute to $F$ apart from the confinement it induces on the
remaining particles which have free energy $F_{_{\cal M}}$:
$F(N,L_0,L)=F_{_{\cal M}}(N,L-L_0)$. Hence, Eq. (\ref{eq:statL0})
also reads
\be
\frac{\partial {\cal R}}{\partial L_0} \,=\, P + N p {\cal L}_p L_0^{p-1}.
\ee
From (\ref{eq:W2}) and (\ref{eq:statW}), we have
${\cal L}_p \propto (\eta_c-\eta_{_{\cal M}})$.
On the other hand, a physically acceptable mixture $\cal M$,
should have a normalizable size distribution, which
imposes $\eta_{_{\cal M}} \leq \eta_c$. Hence, ${\cal L}_p \geq 0$
so that the derivative (\ref{eq:statL0}) cannot vanish, and is always
positive (a situation already encountered in \cite{BC01}).
The minimization with respect to $L_0$ therefore
leads one to choose for $L_0$ the
minimum possible value
compatible with $\eta_{_{\cal M}} \leq \eta_c$.
The optimal value of $L_0$ is thus
0 for $\eta<\eta_c$, and such that
$\eta_{_{\cal M}} = \eta_c$ whenever $\eta>\eta_c$. In other
words
\be
\eta_0 = \left\{
\begin{array}{c l}
 0 &\hbox{ for } \eta<\eta_c\\
\displaystyle\frac{\eta-\eta_c}{1-\eta_c} & \hbox{ for } \eta>\eta_c.
\end{array}
\right.
\ee

\end{appendix}


\vspace{0.3cm}


\vspace{0.2cm}

\begin{thebibliography}{00}

\bibitem{EH05}  M. R. Evans and T. Hanney, J. Phys. A {\bf 38}, R195 (2005)

\bibitem{maj09} S.N. Majumdar, Les Houches (2008) lecture notes,
arXiv:0904:4097

\bibitem{OEC}  O.J.~O'Loan, M.R.~Evans and M.E.~Cates, Phys. Rev. E {\bf
58}%
, 1404 (1998)

\bibitem{KMH}
J. Kaupuzs, R. Mahnke, R.J. Harris
Phys. Rev. E {\bf 72}, 056125 (2005)

\bibitem{granular}  
D. van der Meer, K. van der Weele, P. Reimann and D. Lohse
J. Stat. Mech.: Theor. Exp., P07021  (2007); J. Torok.
Physica A {\bf 355} 374 (2005).

\bibitem{KLMST}  Y.~Kafri, E.~Levine, D.~Mukamel, G.M.~Sch\"{u}tz and
J.~T{%
\"{o}}r{\"{o}}k, Phys.~Rev.~Lett. {\bf 89}, 035702 (2002).

\bibitem{AELM}
A.G.~Angel, M.R.~Evans, E.~Levine and D.~Mukamel,
Phys. Rev. E {\bf 72}, 046132 (2005)

\bibitem{AHE06}
A. G. Angel, T. Hanney, and M. R. Evans,
Phys. Rev. E {\bf 73}, 016105 (2006)


\bibitem{CLMNW}  S.N. Coppersmith, C.-h. Liu, S. Majumdar, O. Narayan, T.A.
Witten Phys. Rev. E., {\bf 53}, 4673 (1996).

\bibitem{MKB}  S.N.~Majumdar, S.~Krishnamurthy and M.~Barma, Phys. Rev.
Lett. {\bf 81}, 3691 (1998); J. Stat. Phys. {\bf 99}, 1 (2000).

\bibitem{RM} R. Rajesh and S.N. Majumdar, Phys. Rev. E {\bf 63}, 036114
(2001).

\bibitem{Bertin}
E. Bertin,
J. Phys. A: Math. Gen. {\bf 39}, 1539 (2006)

\bibitem{BBJ}  P. Bialas, Z. Burda, and D. Johnston, Nucl. Phys. B {\bf
493}%
, 505 (1997).

\bibitem{Evans00}  M. R.~Evans, Braz. J. Phys. {\bf 30}, 42 (2000).

\bibitem{MEZ05}  S.N. Majumdar, M. R. Evans, and  R. K. P. Zia,
Phys. Rev. Lett. {\bf 94}, 180601 (2005)

\bibitem{EMZ06} M. R. Evans, S.N. Majumdar, and R. K. P. Zia,
J. Stat. Phys. {\bf 123} 357 (2006)

\bibitem{BC01}
R. Blaak and J. A. Cuesta,
J. Chem. Phys. {\bf 115}, 963 (2001)

\bibitem{CS02}
J. A. Cuesta and R. P. Sear, Europhys. Lett. {\bf 55}, 451 (2001) ;
Phys. Rev. E {\bf 65}, 031406l (2002)        

\bibitem{ZBTCF99}
J. Zhang, R. Blaak, E. Trizac, J. A. Cuesta, and D. Frenkel,
J. Chem. Phys. {\bf 110}, 5318 (1999)

\bibitem{EH03}
M. R. Evans, T. Hanney,
 J. Phys. A: Math. Gen. {\bf 36} (2003) L441-L447

\bibitem{Tonks}
L. Tonks,
Phys. Rev. {\bf 50}, 955 (1936).


\bibitem{EMZ04}  M. R. Evans, S.N. Majumdar, R. K. P. Zia,
J. Phys. A: Math. Gen {\bf 37} (2004) L275





\bibitem{TP08}
E. Trizac and I. Pagonabarraga,
Am. J. Phys. {\bf 76}, 777 (2008).


\bibitem{Dragu}
A. Dragulescu and V. Yakovenko, Physica A {\bf 299}, 213 (2001);
Eur. Phys. J. B {\bf 17}, 723 (2001).

\bibitem{B00}
R. Blaak,
J. Chem. Phys. {\bf 112}, 9041 (2000).

\bibitem{SS82}
J.J. Salacuse and G. Stell,
J. Chem. Phys. {\bf 77}, 3714 (1982).

\bibitem{TheOtherEvans}
see e.g R. Evans, in ``Fundamentals of Inhomogeneous Fluids'', edited by
D. Henderson
(Marcel Dekker, New York, 1992).











\end{thebibliography}
\end{document}